\documentclass[aps, prl, singlecolumn, superscriptaddress, letterpaper]{revtex4-2}

\usepackage{graphicx} 
\usepackage{amssymb}
\usepackage{amsmath}
\usepackage{float}
\usepackage{txfonts}
\usepackage{bm}
\usepackage{color}
\usepackage[colorlinks=true, linkcolor=blue, anchorcolor=black, citecolor=blue, filecolor=black, menucolor=black, runcolor=black, urlcolor=blue]{hyperref}
\usepackage{mathrsfs}

\usepackage[normalem]{ulem} 

\begin{document}

\title{Photonic bilayer Chern insulator with corner states}

\author{Subhaskar Mandal}
\thanks{These authors contributed equally.}
\affiliation{Division of Physics and Applied Physics, School of Physical and Mathematical Sciences, Nanyang Technological University,
Singapore 637371, Singapore}

\author{Ziyao Wang}
\thanks{These authors contributed equally.}
\affiliation{State Key Laboratory of Optical Fiber and Cable Manufacturing Technology, Department of Electronic and Electrical Engineering, Guangdong Key Laboratory of Integrated Optoelectronics Intellisense, Southern University of Science and Technology, Shenzhen 518055, China}

\author{Rimi Banerjee}
\affiliation{Division of Physics and Applied Physics, School of Physical and Mathematical Sciences, Nanyang Technological University,
Singapore 637371, Singapore}

\author{Hau Tian Teo}
\affiliation{Division of Physics and Applied Physics, School of Physical and Mathematical Sciences, Nanyang Technological University,
Singapore 637371, Singapore}

\author{Peiheng Zhou}
\affiliation{National Engineering Research Center of Electromagnetic Radiation Control Materials, State Key Laboratory of Electronic Thin Film and Integrated Devices, University of Electronic Science and Technology of China, Chengdu 610054, China}

\author{Xiang Xi}
\email{xix@sustech.edu.cn}
\affiliation{State Key Laboratory of Optical Fiber and Cable Manufacturing Technology, Department of Electronic and Electrical Engineering, Guangdong Key Laboratory of Integrated Optoelectronics Intellisense, Southern University of Science and Technology, Shenzhen 518055, China}

\author{Zhen Gao}
\email{gaoz@sustech.edu.cn}
\affiliation{State Key Laboratory of Optical Fiber and Cable Manufacturing Technology, Department of Electronic and Electrical Engineering, Guangdong Key Laboratory of Integrated Optoelectronics Intellisense, Southern University of Science and Technology, Shenzhen 518055, China}

\author{Gui-Geng Liu}
\email{guigeng001@e.ntu.edu.sg}
\affiliation{Division of Physics and Applied Physics, School of Physical and Mathematical Sciences, Nanyang Technological University,
Singapore 637371, Singapore}

\author{Baile Zhang}
\email{blzhang@ntu.edu.sg}
\affiliation{Division of Physics and Applied Physics, School of Physical and Mathematical Sciences, Nanyang Technological University,
Singapore 637371, Singapore}
\affiliation{Centre for Disruptive Photonic Technologies, Nanyang Technological University, Singapore 637371, Singapore}

\maketitle
{\textbf{Photonic Chern insulators can be implemented in gyromagnetic photonic crystals with broken time-reversal (TR) symmetry. They exhibit gapless chiral edge states (CESs), enabling unidirectional propagation and demonstrating exceptional resilience to localization even in the presence of defects or disorders. However, when two Chern insulators with opposite Chern numbers are stacked together, this one-way nature can be nullified, causing the originally gapless CESs to become gapped. Recent theoretical works have proposed achieving such a topological phase transition in condensed matter systems using antiferromagnetic thin films such as MnBi$_2$Te$_4$  or by coupling two quantum spin/anomalous Hall insulators, but these approaches have yet to be realized experimentally. In a bilayer gyromagnetic photonic crystal arranged in an antiferromagnetic layer configuration, our experimental observations reveal that interlayer coupling initiates a transition from a Chern insulating phase to a higher-order  topological phase. This transition results in the gapping of CESs and triggers the emergence of corner states within the bandgap. The corner mode energy within the gap can be attributed to CESs interaction, forming a Jackiw-Rebbi topological domain wall mode at the corner. These states exhibit heightened resilience against defects, setting them apart from their time-reversal symmetric counterparts.}}

\begin{figure}[t]
\centering
\includegraphics[width = 0.75 \textwidth]{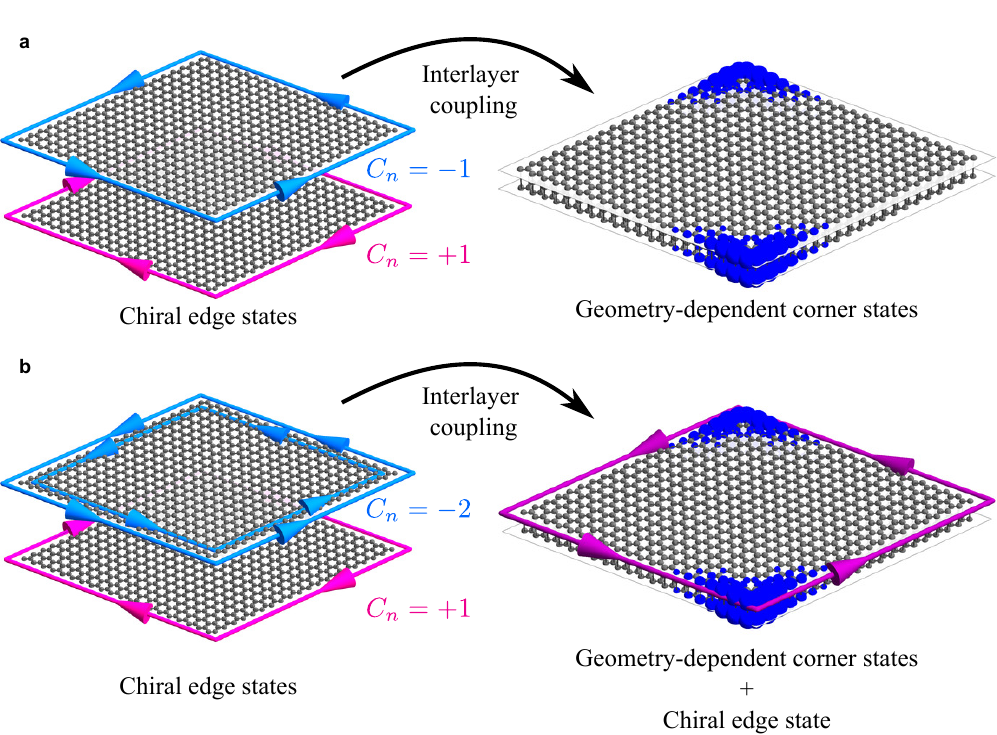}%
\caption{\textbf {Chern insulator to HOTI transition. a,} Upon coupling the two layers, a bilayer CI characterized by Chern numbers of opposite signs ($C_n=\pm1$) transforms into a HOTI, featuring in-gap topological corner modes. \textbf {b,} Coupled CIs having Chern numbers of unequal and opposite signs  ($C_n=+1,~-2$) can host both CES and corner modes within the same bulk bandgap.}
\label{Fig1}
\end{figure}

\vspace{15pt}
\noindent \textbf{\large{Introduction}}

Chern insulators (CIs), reliant on the breaking of time reversal (TR) symmetry, have remained a fundamental concept in the field of topological physics. Within a two-dimensional (2D) Chern insulator, there exist one-dimensional (1D) gapless chiral edge states (CESs) along its boundary, showcasing a phenomenon intricately governed by the principles of bulk-edge correspondence \cite{RevModPhys.82.3045}. These chiral edge states are celebrated for their extraordinary robustness as unidirectional transport channels, remaining impervious to localization caused by defects or disorder, as long as the Chern number $C_n$ characterizing the bulk topology remains unaltered \cite{Lu2014,Khanikaev2017,RevModPhys.91.015006}. 

Nonetheless, this one-way propagation can be countered by stacking two Chern insulators with Chern numbers of opposite signs in a bilayer system. In such a configuration, the interaction between the CIs induces the gapping of the originally gapless CESs. In scenarios adhering to specific crystalline symmetries, these CESs can be gapped across the entire system, with the notable exception of the corners. This phenomenon gives rise to a higher-order topological insulator (HOTI), driven by the interaction between the Chern insulators. Recent theoretical studies suggest that such a topological phase transition in condensed matter systems could be achieved using bilayers of antiferromagnetic thin films like MnBi$_2$Te$_4$ \cite{zhan2023design} or or by coupling two quantum spin/anomalous Hall insulators  \cite{liu2024interlayer}. However, these methods have not yet been demonstrated experimentally.

HOTIs have emerged as a captivating phase of matter, characterized by the emergence of robust boundary states with dimensions (D-d) in dth order D-dimensional finite systems \cite{doi:10.1126/science.aah6442,PhysRevB.96.245115,Xie2021}. In 2D HOTIs, 0D states are precisely localized at the corners, igniting substantial interest in photonics \cite{Peterson2018,Noh2018,Mittal2019,PhysRevLett.125.213901,Xie2020,Schulz2022}. This interest stems from their potential applications, including high-quality cavities \cite{Ota:19}, low-threshold lasers \cite{Zhang2020}, multipolar lasers \cite{Kim2020}, and exciton-polariton lasers \cite{doi:10.1126/sciadv.adg4322,bennenhei2024organic}. The topological protection of HOTIs primarily relies on bulk polarization or quadrupole moments governed by the system's crystalline symmetries \cite{doi:10.1126/science.aah6442,PhysRevB.96.245115}. While these symmetries ensure mode degeneracy, they do not guarantee energy confinement within the bulk bandgap \cite{PhysRevB.96.245115,doi:10.1126/science.aba7604}. The presence of modes within the bandgap is influenced by local symmetries like chiral and particle-hole symmetry, which are challenging to externally control and often disrupted in practical photonic systems. This disruption results in topological modes coexisting with bulk states, making them more susceptible to defects and disorders, compromising their robustness and stability. In contrast, photonic CESs consistently reside within the bulk bandgap without maintaining any additional symmetries other than the TR symmetry breaking. Therefore, any corner state resulting from the interaction of CESs in the bilayer configuration will always exist within the bulk bandgap, provided that the Chern number of each individual layer remains non-trivial, making them more robust to their time-reversal symmetric counterparts. 

The gyromagnetic photonic crystal has emerged as an exceptional testing ground for investigating unparalleled TR-broken topological effects in photonics, including CESs \cite{Wang2009,PhysRevLett.106.093903}, non-Hermitian CESs \cite{PhysRevLett.132.113802}, antichiral states \cite{PhysRevLett.125.263603,Jianfeng_Chen_2022_Opto_Electronic_Science,Xi2023}, topological Anderson insulators \cite{PhysRevLett.125.133603}, and topological Chern vectors \cite{Liu2022}, among others. Notably, the realization of HOTIs within gyromagnetic photonic crystals has been achieved recently \cite{PhysRevB.107.014105,10.1093/nsr/nwae121}. However, the significance of CESs in these studies was overlooked, opting instead for protection through bulk polarization and quadrupole moment. These mechanisms fail to safeguard energy and may lead to the emergence of corner modes intertwined with the bulk continuum as shown in the latter scenario.

In this work, we predict theoretically and  verify experimentally the realization of a HOTI by employing a pair of coupled Chern insulators with Chern numbers of opposite signs (see Fig.~\ref{Fig1}). To achieve this, we utilize a bilayer gyromagnetic photonic crystal in the antiferomagnetic layer configuration at the microwave scale. By manipulating the magnetization orientation in the two layers of the bilayer structure, we successfully create Chern insulators with opposite Chern numbers. Through intentional coupling between the layers, the system undergoes a transition from a Chern insulator phase to a higher-order topological phase. This transition induces a gap in the CESs and gives rise to the emergence of in-gap corner modes, as evidenced by transmission measurements and near-field scanning. Notably, not all corners can host the topological corner modes, making the corner modes obtained in this scheme geometry-dependent. The origin of these corner modes can be explained using the Jackiw-Rebbi topological domain wall mode \cite{PhysRevD.13.3398}. Expanding upon our approach, we demonstrate that corner modes can be effectively induced by coupling Chern insulators featuring distinct bandgap or by coupling the CES with the flat zigzag edge states in graphene. Additionally, our exploration uncovers an innovative topological phase achieved by combining Chern insulators characterized by unequal Chern numbers. Within this phase, we observe the coexistence of gapless CES and topological corner modes within the same bulk bandgap, which has not been reported previously to the best of our knowledge. The corner modes obtained in our scheme exhibit exceptional robustness against defects, setting them apart from their counterparts in time-reversal symmetric systems.

\begin{figure*}[t]
\centering
\includegraphics[width = 0.95 \textwidth]{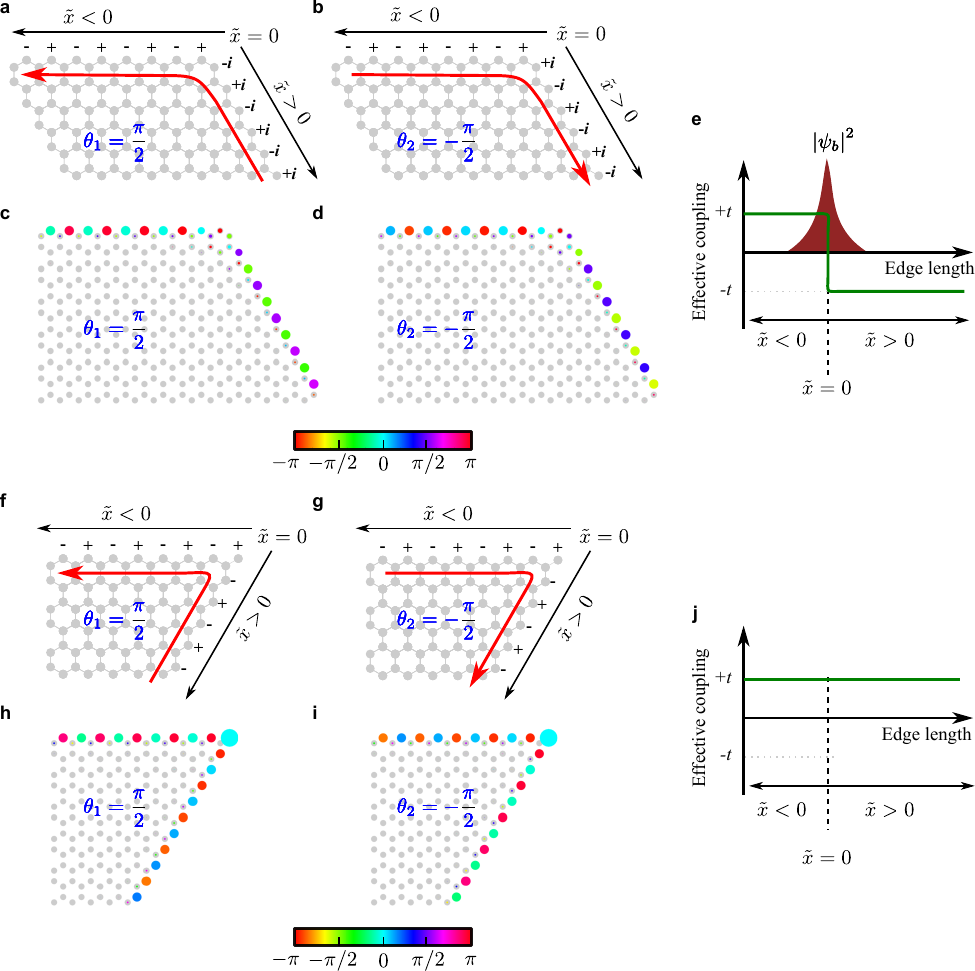}%
\caption{\textbf {Jackiw-Rebbi formalism of the corner modes.} \textbf{a-b}, Phases of CESs in the absence of inter-layer coupling. Following a $120^\circ$ corner, a $\pi$ phase disparity arises between the CESs in the two layers having $\theta_1=+\pi/2$ and $\theta_2=-\pi/2$, respectively. \textbf{c-d}, Computational evaluation of CES phases utilizing the tight-binding model, illustrating the $\pi$-phase difference subsequent to a $120^\circ$ corner. \textbf{e}, The phase alignment prerequisite necessitates an inverse sign for the effective coupling $t$ before and after the $120^\circ$ corner, culminating in the emergence of a zero-energy state with exponential localization at the corner. \textbf{f-j}, In scenarios involving a $60^\circ$ corner, the CESs lack a discernible phase difference prior to and after the corner. Consequently, the effective coupling $t$ remains unaltered in sign, precluding the manifestation of an exponentially localized zero-energy state at the corner.}
\label{FigTBJR}
\end{figure*}

\begin{figure*}[t]
\centering
\includegraphics[width =0.9\textwidth]{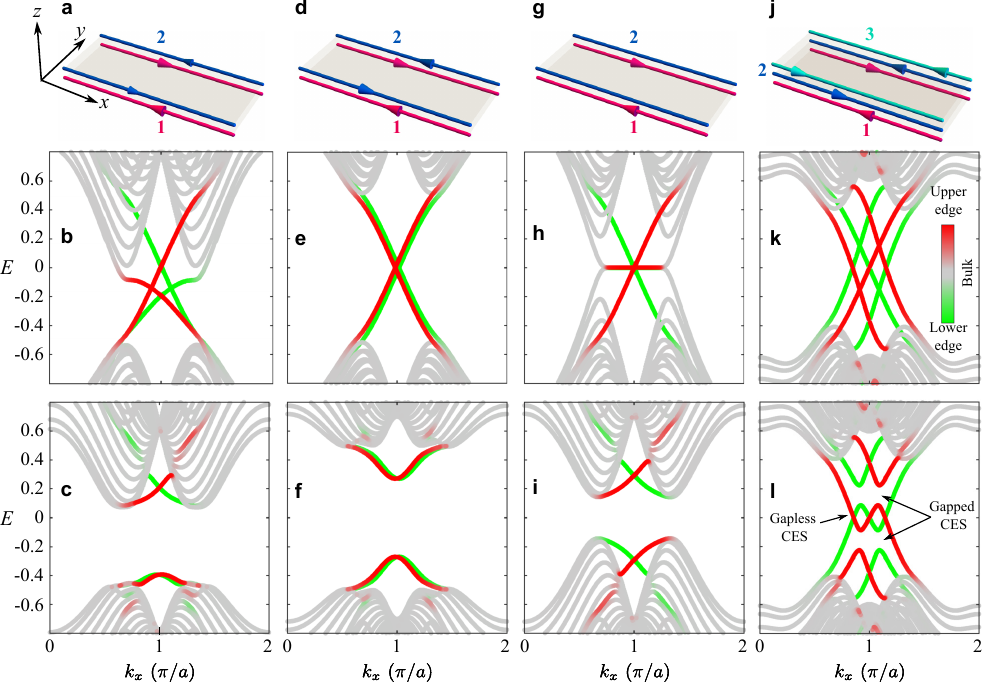}%
\caption{\textbf {Various CESs Interaction Scenarios.} \textbf {First row:} Schematics depicting different system configurations under investigation. \textbf {a-c,} CIs with distinct and opposite NNN coupling phases, where $\theta_1 = \pi/2$ and $\theta_2 = -\pi/8$.  \textbf {d-f}, CIs with same but opposite NNN coupling phases, where $\theta_1=\pi/2$ and $\theta_2=-\pi/2$. \textbf {g-i}, A CI coupled with graphene, where $\theta_1 = \pi/2$ and $J_{nn} = 0$ in layer 2. \textbf{j-l}, A trilayer system demonstrating interactions between CIs with Chern numbers $C_n = +1$ and $C_n = -2$. Here, $\theta_1 = \pi/2$, $\theta_2 = \theta_3 = -\pi/2$, $J_{nn} = 0.3$ in layer 3 with $J_{23} = 0.3$. \textbf {Second row:}  Strip band structure of the system without interlayer coupling ($J_{12} = 0$). \textbf {Third row:} Strip band structure of the system with interlayer coupling, illustrating the gapping of the CESs. Here, $J_{12} = 0.1$ for the last column and $J_{12} = 0.3$ for the others. $J_{nn} = 0.1$ where not specified, while $J_n = 1$.}
\label{Fig2TB}
\end{figure*}

\vspace{15pt}
\noindent \textbf{\large{Theoretical scheme}}
\vspace{5pt}

\noindent\textbf{Tight-binding model}

To illustrate our approach, we consider the Haldane model \cite{PhysRevLett.61.2015} as an example of a CI. The Hamiltonian of this model can be expressed in the sub-lattice basis as:
\begin{align}\label{HalMod}
H_S(\mathbf{k},\theta)&= \begin{bmatrix}
\alpha(\mathbf{k},\theta) &\beta(\mathbf{k})\\
 \beta^*(\mathbf{k}) &\alpha(\mathbf{k},-\theta)
\end{bmatrix},
\end{align}
where the matrix elements $\alpha$ and $\beta$ are given by:
\begin{align}
\alpha(\mathbf{k},\theta)=&2J_{nn}\left[\cos(\sqrt{3}bk_x-\theta)+2\cos\left(\frac{3bk_y}{2}\right)\cos\left(\frac{\sqrt{3}bk_x}{2}+\theta\right)\right],\notag\\
\beta(\mathbf{k})=&J_n\left[1+2\cos\left(\frac{\sqrt{3}bk_x}{2}\right)\exp\left(i\frac{3bk_y}{2}\right)\right].\notag
\end{align}
Here, $b$ is the lattice constant, $J_n$ is the nearest neighbor coupling, and $J_{nn}$ is the next nearest neighbor (NNN) coupling with phase $\theta$. It is well known that the above Hamiltonian in Eq.~\ref{HalMod} represents a CI for $J_{nn}\neq 0$ and $\theta\neq 0,\pi$ \cite{PhysRevLett.61.2015}.

Moving forward, we proceed to construct a coupled bilayer system, as depicted in Fig.~\ref{Fig1}. In this arrangement, each layer is represented by the Hamiltonian described in Eq.~\ref{HalMod}. The resulting Hamiltonian for this bilayer system can be expressed as follows:
\begin{align}
H_B(\mathbf{k})&=
\begin{bmatrix}
H_S(\mathbf{k},\theta_1) & J_{12} \mathbf{I}_{2\times 2}\\
 J_{12}  \mathbf{I}_{2\times 2} & H_S(\mathbf{k},\theta_2)
\end{bmatrix}.
\end{align}

In this expression, $\theta_1$ and $\theta_2$ denote the phases of the NNN couplings in the two layers, $J_{12}$ represents the interlayer coupling, and $\mathbf{I}_{2\times 2}$ denotes an identity matrix. Notably, when $\text{sgn}(\theta_1)/\text{sgn}(\theta_2)=-1$, the two layers of $H_B$ correspond to CIs with Chern numbers of opposite signs $\pm C_n$ and host CESs with opposite chirality. Our primary interest lies in understanding the consequences of CESs interaction at the corner of the sample.

To investigate this, we shift our focus to the edges of the sample, where the effective Hamiltonian near the crossing energy of the CESs can be expressed as:
\begin{align}\label{EdgeHam}
H_e= -i\frac{\partial}{\partial \tilde{x}}\tau_z+t\cos\zeta_{1,2}\tau_x+t\sin\zeta_{1,2}\tau_y.
\end{align} 
Here, $\tilde{x}$ denotes the coordinate along the edges of a finite sample involving a corner, $\tau$ is the Pauli matrix representing the layers. The first term represents the counter-propagating CESs in the two layers, whereas the last two terms represent the coupling between the CESs. $t$ stands for the amplitude of the coupling, and $\zeta_{1,2}$ represent the phases of the couplings before and after the corner. Notably, $\zeta_{1,2}$ can assume any value in our analysis and due to the presence of the corner, $\zeta_{1}$ may not be the same as $\zeta_{2}$. Consequently, we can identify a Jackiw-Rebbi  \cite{PhysRevD.13.3398} type exponentially localized topological bound state at $\tilde{x}=0$, which takes the form (see Supplementary Information I):
\begin{align}
\psi_b(\tilde{x}<0)=Ce^{\tilde{x}\sqrt{t^2-\varepsilon^2}}\left[
\begin{matrix}
e^{-i\zeta_1}\left(\varepsilon-i\sqrt{t^2-\varepsilon^2}\right)/t\\
1
\end{matrix}
\right],\label{Bound1}\\
\psi_b(\tilde{x}>0)=Ce^{-\tilde{x}\sqrt{t^2-\varepsilon^2}}\left[
\begin{matrix}
e^{-i\zeta_2}\left(\varepsilon+i\sqrt{t^2-\varepsilon^2}\right)/t\\
1
\end{matrix}
\right].\label{Bound2}
\end{align}
Here, $C$ is the normalization constant, and $\varepsilon$ is the  bound state energy. This bound state can be considered as the manifestation of higher-order topological corner modes resulting from the interaction of the CESs. By ensuring the continuity of Eqs.~\ref{Bound1}-\ref{Bound2} at $\tilde{x}=0$, we can determine  $\varepsilon$, which is given by:
\begin{align} \label{Bound_energy}
\varepsilon=t\cos\left[\left(\zeta_1-\zeta_2\right)/2\right].
\end{align}

Now we obtain the Jackiw-Rebbi bound mode $\psi_b$ for $\theta_1=\pi/2$ and  $\theta_2=-\pi/2$ in the two layers, respectively. In this configuration, the CESs intersect at $k_x = \pi/a$ , where 
$a$ is the periodicity along the $x$ direction with zigzag edges. Near this intersection, the eigenstates of the uncoupled system exhibit a periodicity of $2a$. Notably, the eigenstates in the two layers differ by a phase shift of $\pi$ before and after the $120^\circ$ corner (see Figs.~\ref{FigTBJR}a-d). Consequently, to satisfy the phase matching condition of the resultant wave function in presence of the interlayer coupling, the effective coupling $t$ must change sign before and after the  $120^\circ$ corner, resulting $\zeta_1-\zeta_2=\pi$. Substituting it back into Eqs.~\ref{Bound1}-  \ref{Bound_energy}, one can obtain exponentially localized zero-energy ($\varepsilon=0$) bound state at the $120^\circ$ corner. On the other hand, for a $60^\circ$ corner, the eigenstates do not show a phase difference between the two layers, resulting in, $\zeta_{1}=\zeta_{2}$ (see Figs.~\ref{FigTBJR}e-h). Consequently, $60^\circ$ corners do not host topological corner modes, leading to geometry-dependent corner modes in our scheme.

\vspace{5pt}
\noindent\textbf{Topological corner modes}

Having anticipated the emergence of corner modes using the Jackiw-Rebbi model, we now study the CES interaction from the projected band structures for various configurations in Fig.~\ref{Fig2TB}. Initially, we examine a generic scenario where the CESs in the two layers propagate counter to each other and are not identical but share a common bandgap ($|\theta_1|\neq|\theta_2|$). In Fig.~\ref{Fig2TB}b, the strip band structure is depicted, with the $x$ direction assumed as periodic and the $y$ direction truncated to feature zigzag edges. The CESs localized at the top and bottom edges of the strip are shown in red and green, respectively, while the bulk bands are shown in grey. In the absence of interlayer coupling, the CESs of the system are gapless. Upon introducing interlayer coupling, the CESs becomes gapped, indicating a phase transition from first-order to higher-order topology (see Fig.~\ref{Fig2TB}c). Due to the broken chiral symmetry, the bands are asymmetric about $E=0$. Figs.~\ref{Fig2TB}d-f explore a scenario where both layers are symmetric with opposite Chern numbers ($C_n$) corresponding to $\theta_1=-\theta_2=\pi/2$.  As shown in Fig.\ref{FigTBJR}, without interlayer coupling, the CESs cross each other at $k_x=\pi/a$, making them periodic with periodicity $2a$. Due to the interlayer coupling  the gapless CESs become gapped. However, in this case, chiral symmetry is preserved, resulting in bands that are symmetric about $E=0$ and maximizing the bandgap. 

To extend the scope, we investigate two additional scenarios. The first involves the interaction between a CI and a graphene layer with $J_{nn}=0$. The interaction induces a bandgap in the edge spectrum (see Figs.~\ref{Fig2TB}g-i). Notably, the overall bandgap is smaller compared to the previous cases, attributed to the absence of a bulk bandgap in the graphene layer.

Lastly, we present a unique topological phase where gapped and gapless CESs coexist, indicating the presence of first-order and higher-order topology within the same bulk bandgap. This is achieved through the interaction between CIs with $C_n=+1$ and $C_n=-2$. The CI with $C_n=-2$ is prepared by coupling two layers with the same NNN coupling phase ($\theta_2=\theta_3=-\pi/2$) and then coupling the effective $C_n=-2$ system to a third layer with the opposite NNN coupling phase  ($\theta_1=\pi/2$). Fig.~\ref{Fig2TB}k illustrates the band structure without interlayer coupling $J_{12}$, showcasing one propagating CES alongside two counter-propagating CESs. The CESs on the same edge cross at two distinct energies. Interlayer coupling $J_{12}$ induces bandgaps at both the energy crossings, rendering two CESs gapped. However, one CES remains gapless (see Fig.~\ref{Fig2TB}l).

\begin{figure*}[t]
\centering
\includegraphics[width =0.9\textwidth]{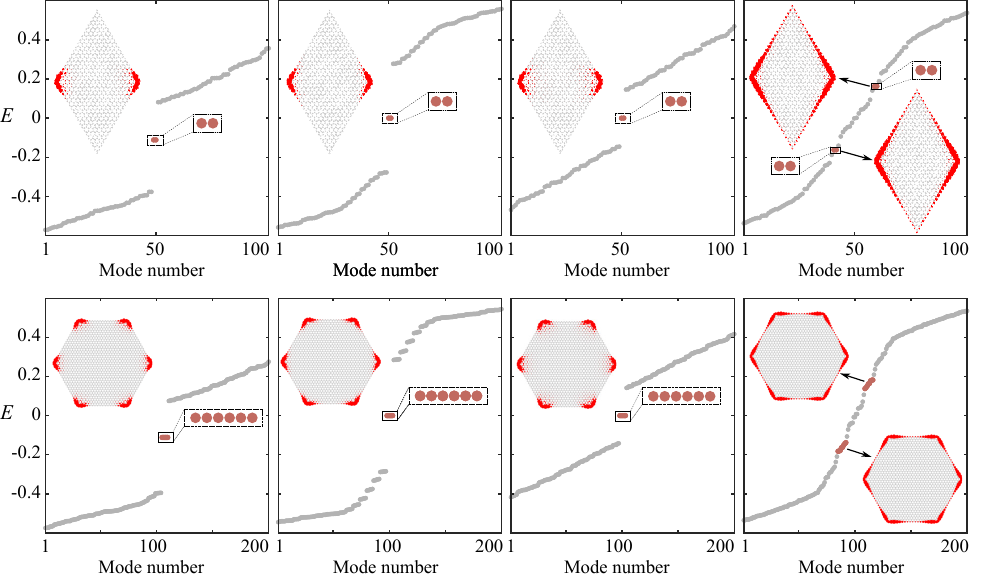}%
\caption{\textbf {Topological corner modes in different geometries.} Different columns correspond to those in Fig.~\ref{Fig2TB}. Eigenfrequencies for rhombic and hexagonal finite structures are shown in the first and second rows, respectively, revealing the emergence of topological corner modes. The insets provide spatial profiles of these corner modes. In the second column, the $E=0$ energy of the corner modes aligns well with the bound state energy obtained in Eq.~\ref{Bound_energy}. In the last column, the corner modes appear at two different gapped CES energies, but they hybridize with the gapless CES. In the first row the corner modes are not located at all four corners, which is in agreement with Fig.~\ref{FigTBJR}.}
\label{Fig_ENTB}
\end{figure*}

Finally, in Fig.~\ref{Fig_ENTB}, we present the topological corner modes in rhombic and hexagonal shaped finite geometries, which appear inside the gapped CESs. For the rhombic case, as predicted by the Jackiw-Rebbi model in Fig.~\ref{FigTBJR}, only $120^\circ$ corners host the topological corner modes, while corner modes are absent at the $60^\circ$ corners. On the other hand, in the hexagonal geometry, all six corners are $120^\circ$. Consequently, six topologically protected corner modes appear at the six corners. The corner modes are pinned to $E=0$ energy for the second column, which is in agreement with the Jackiw-Rebbi bound state energy obtained from Eq.~\ref{Bound_energy}. For the last column, which represents the CES interaction having $C_n=+1$ and $-2$, corner modes at both CESs gapping energies appear (see Fig.~\ref{Fig2TB}l). The energy spectrum of the finite system lacks a bandgap due to the presence of the gapless CES. The corner modes in this case hybridize with the CES, making them less localized at the corners compared to the previous cases.

\begin{figure*}[t]
\centering
\includegraphics[width = 1 \textwidth]{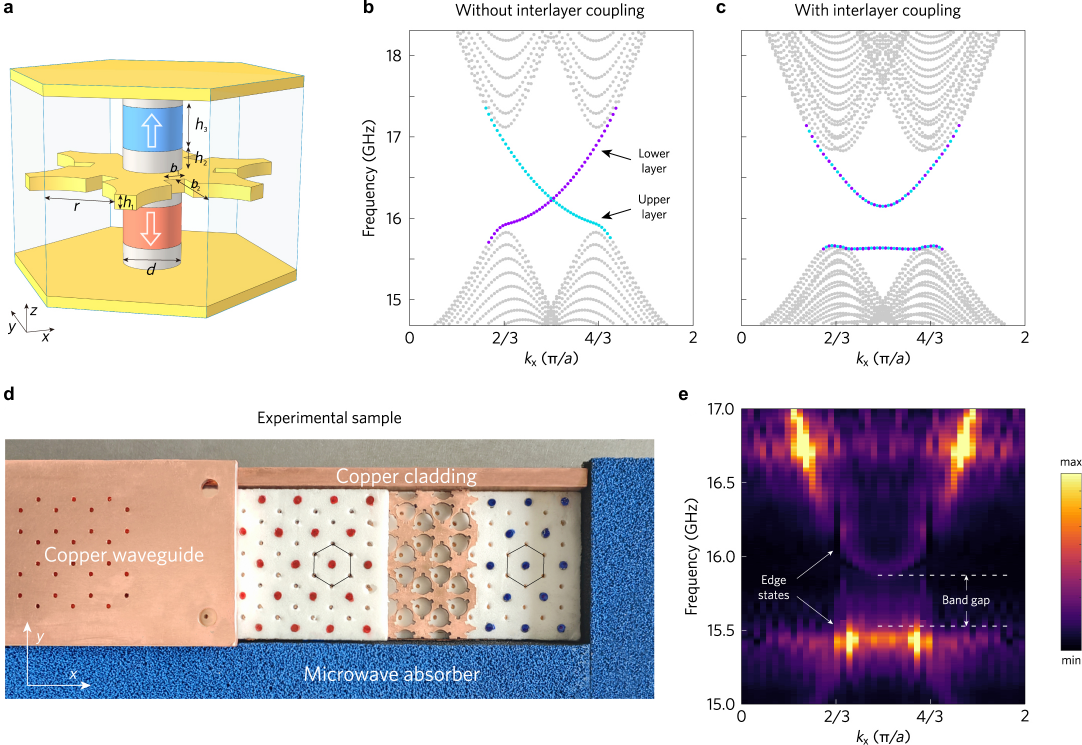}%
\caption{\textbf{Design and band structure of a bilayer gyromagnetic photonic crystal.} \textbf{a}, Schematic of the bilayer unit cell, comprising gyromagnetic cylinders (blue and red) with a diameter of $d = 2.8$ mm and a height $h_3 = 2$ mm. The cylinders are magnetically biased along the $z$-direction using permanent magnets (grey cylinders). The unit cell length along the $x$-direction is $a = 12$ mm. \textbf{b-c}, Numerically calculated band structure of a strip composed of 23 unit cells with PEC boundary conditions along the $y$-direction and periodic boundary conditions along the $x$-direction. The bilayer gyromagnetic photonic crystal exhibits gapless chiral edge states along one PEC boundary without interlayer coupling \textbf{b}, and the chiral edge states become gapped with interlayer coupling \textbf{c}. 
\textbf{d}, Photograph of a strip sample sandwiched between copper waveguides, with copper cladding on the top edge as PEC boundary while other edges are covered with microwave absorber. \textbf{e}, Measured band structure showing the presence of gapped edge states, consistent with the numerical results in \textbf{c}.}
\label{Fig2}
\end{figure*}

\begin{figure*}[t]
\centering
\includegraphics[width = 0.95 \textwidth]{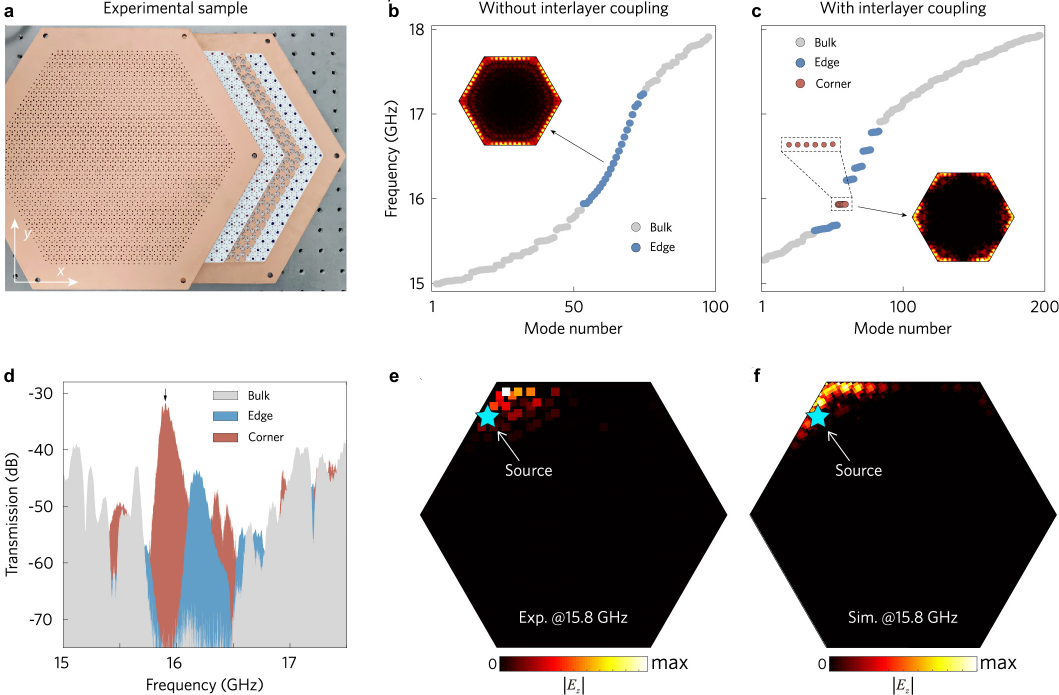}%
\caption{\textbf{Higher-order topological corner states in a hexagonal sample}. \textbf{a}, Photograph of a hexagonal sample with copper bars along all six edges as PEC boundaries. \textbf{b-c}, Numerically calculated eigenfrequencies for single layer photonic crystal without interlayer coupling and the BPhC, respectively. Both systems exhibit gapped bulk modes, but only the BPhC displays gapped edge states, with the emergence of six higher-order topological corner modes. The inset in \textbf{c} illustrates the spatial profile of the six higher-order topological corner states, while the inset in \textbf{b} shows one of the chiral edge states. \textbf{d}, Measured transmission spectra from the bulk, edge, and corner of the sample as a function of frequency. \textbf{e}, The measured field distribution corresponding to the corner mode frequency, demonstrating good agreement with the numerically calculated spatial profile shown in \textbf{f}.  }
\label{Fig3}
\end{figure*}

\begin{figure*}[t]
\centering
\includegraphics[width = 0.75 \textwidth]{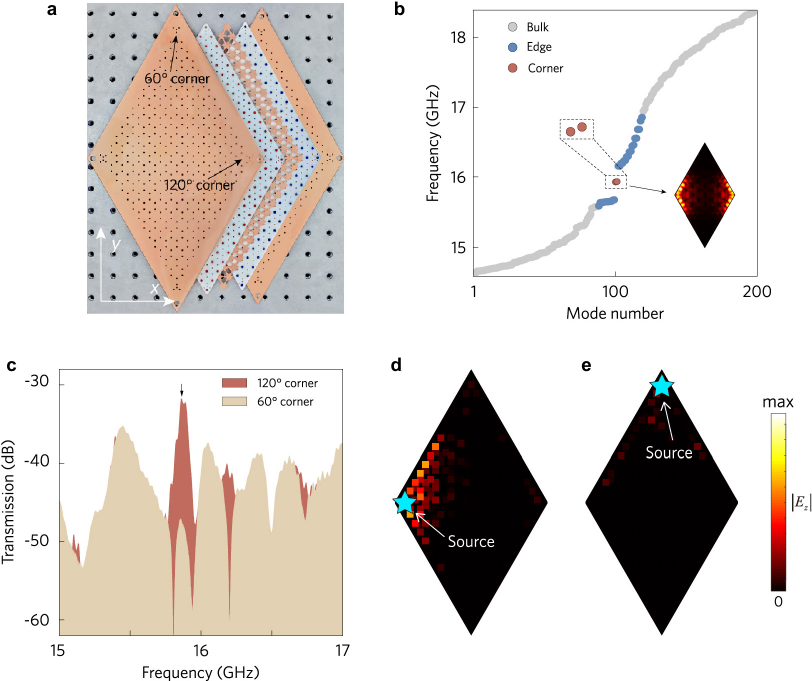}
\caption{\textbf{Geometry-dependent higher-order topological corner modes}. \textbf{a}, Photograph of the rhombic sample with copper bars covering the edges to realize PEC boundaries. \textbf{b}, Numerically calculated eigenfrequencies of the structure depicted in \textbf{a}, revealing the existence of two corner modes at the $120^\circ$ corners rather the $60^\circ$ corners. Inset shows the electric field profile of the corner modes. \textbf{c}, Measured transmission spectrum of topological corner modes at the $120^\circ$  and $60^\circ$  corners, respectively. \textbf{d-e}, Measured electric field distributions of the topological corner mode at the $120^\circ$ \textbf{d} and $60^\circ$ \textbf{e}, corners, respectively, with the position of the point source indicated with cyan star. For the purpose of comparison, consistent colorbar limits have been applied to both \textbf{d} and \textbf{e}.}
\label{Fig5}
\end{figure*}

\vspace{15pt}
\noindent \textbf{\large{Experimental results}}
\vspace{5pt}

\noindent\textbf{Strip band structure}

We experimentally demonstrate the existence of higher-order topological corner modes within a microwave magnetic photonic system. These systems are well-known for studying effects related to the Haldane model, such as CIs \cite{Wang2009}, antichiral edge states \cite{PhysRevLett.125.263603}, and more. We specifically selected a configuration where the TR symmetry breaking is identical and opposite in the two layers, strategically maximizing the resultant bandgap. The unitcell of our designed bilayer photonic crystal (BPhC) is shown in Fig.~\ref{Fig2}a, within each layer comprises a gyromagnetic cylinder sandwiched and biased by a pair of permanent magnets. The magnetic field with an average of 220 mT along the $+z~(-z)$ direction are applied to the gyromagnentic rod in the upper (lower) layer. The two layers are isolated by a copper metallic plate. Individually, the upper and lower layers are photonic Chern insulators with Chern numbers $+1$ and $-1$, respectively (see Supplementary Information III). The calculated band dispersion of the edge state is shown in Fig.~\ref{Fig2}b, which hosts edge states with opposite chirality in the two layers. Then the interlayer coupling is introduced by drilling patterned air holes onto the isolated copper plate. The interlayer coupling in the bilayer Chern insulator results in a gap of the chiral edge state, as illustrated in Fig.~\ref{Fig2}c. The system exhibits a Wilson loop winding reminiscent of Chern insulators with $C_n=\pm1$, indicative of Wannier obstruction. Nonetheless, this winding can be trivialized by incorporating additional lower bands lacking topological characteristics (see Supplementary Information III).

We constructed a strip sample consisting of $60\times 5$ unit cells, as depicted in Fig.~\ref{Fig2}d. The upper edge of the sample is equipped with a copper cladding to establish a perfect electric conductor (PEC) boundary, effectively preventing the leakage of electromagnetic waves into the surroundings. Microwave absorbers are applied to the remaining edges of the sample. The magnets and YIGs are securely positioned using dielectric foam. Air holes are strategically drilled on the top waveguide plate to facilitate excitation and measurement. By introducing a source at the upper edge, we can stimulate the edge states and obtain their real-space profile. Subsequently, a Fourier transform can be applied to reveal the dispersion of the edge states. Fig.~\ref{Fig2}e displays the dispersion of the edge states, which is in good agreement with the numerically calculated band structure, confirming the presence of a gap of around $0.5$ GHz in the edge states.

\vspace{15pt}
\noindent\textbf{Demonstration of topological corner modes}

To verify the appearance of the topological corner modes, we construct a hexagonal sample having $13$ unit cells in each arms. A photograph of the sample is shown in Fig.~\ref{Fig3}a. Similar to the previous figure, PEC boundaries are realized at all the six edges of the sample using copper cladding. We numerically calculate the eigenfrequencies of our BPhC, as illustrated in Fig.~\ref{Fig3}c. Consistent with the band structure depicted in Fig.~\ref{Fig2}c, the finite sample exhibits gapped bulk bands and the emergence of gapped edge states. Notably, we obtain six topologically protected higher-order corner modes within the gapped edge modes, in agreement with the tight-binding model. The summation of the spatial profiles of the $z$-component of the electric field, $\sum{_{n=1}^6\left|E_z^n\right|}$, for the six corner modes is displayed in the inset of Fig.~\ref{Fig3}c. In contrast, for a single layer system, unlike the BPhC, the chiral edge states span the entire bulk bandgap, as depicted in Fig.~\ref{Fig3}b. The spatial profile of one of the chiral edge states, shown in the inset of Fig.~\ref{Fig3}b, clearly demonstrates that the energy is localized at the edges of the sample rather than the corners. In order to experimentally characterize the topological corner modes, we conducted transmission measurements with the source and probe positioned at various locations in our hexagonal sample: the bulk, edge, and near one corner. The measured bulk and edge transmissions, as depicted in Fig.~\ref{Fig3}d, clearly exhibit a dip around $15.8$ GHz, indicating the presence of both bulk and edge bandgaps. Notably, the transmission curve for the corner peaks near $15.8$ GHz, providing evidence for the existence of topological corner modes in close proximity to that frequency. To further confirm the presence of the corner mode, we directly scanned the field distribution, as shown in Fig.~\ref{Fig3}e, revealing a pronounced concentration of the field at the corner, which shows good agreement with the numerically simulated case as shown in Fig.~\ref{Fig3}f.

\vspace{15pt}
\noindent\textbf{Geometry dependent topological corner modes}

Next, we engineered a sample with a rhombic geometry, as depicted in Fig.~\ref{Fig5}a. Similar to previous structures, the copper cladding encompasses all four edges, establishing PEC boundaries. As predicted from the tight-binding analysis, the numerically calculated eigenfrequencies distinctly unveil the presence of two topologically protected corner modes, as depicted in Fig.~\ref{Fig5}b. The spatial profile of these corner modes is highlighted in the inset of Fig.~\ref{Fig5}b, illustrating their exclusive localization at $120^\circ$ corners, devoid of any contribution from the $60^\circ$corners.

To experimentally validate this phenomenon, we conducted transmission measurements by situating a source near the $120^\circ$ and $60^\circ$ corners. The transmission spectrum is illustrated in Fig.~\ref{Fig5}c, demonstrating a notably higher transmission peak at 15.8 GHz corresponding to the $120^\circ$ corner compared to the $60^\circ$ corner. This discrepancy underscores the existence of solely the $120^\circ$ corner states at that specific frequency. Furthermore, we visualized the field distribution at the two corners, as shown in Fig.~\ref{Fig5}d-e. Evidently, a pronounced field localization is observed at the $120^\circ$ corner, confirming the presence of the topological corner mode. In contrast, the $60^\circ$ corner displays significantly lower field intensity, signifying the absence of the topological corner mode. To accentuate the differentiation between the two types of corners, the colorbar limits in Fig.~\ref{Fig5}d-e have been maintained identically.

In Supplementary Information V, we verify various other layer configurations and numerically realize the additional cases shown in Fig.~\ref{Fig2TB}. We also conducted a comparative analysis of the robustness against defects exhibited by these corner modes, contrasting them with their counterparts in a time-reversal symmetric system (see Supplementary Information VI). Although the Jackiw-Rebbi topological bound modes can explain the existence of the corner modes for various layer configurations, the symmetric layer configuration demonstrated in the experiments respects an additional $C_{2x}$ mirror symmetry, allowing us to define a non-trivial bulk mirror winding number that also protects the corner modes (see Supplementary Information IV).

\vspace{15pt}
\noindent\textbf{\large{Conclusions}}

In summary, our study presents a significant advancement in the field of topological photonics by theoretically predicting and experimentally realizing a new type of higher-order topological insulator. This is achieved by coupling different layers of Chern insulators through interlayer coupling. Using this approach, we discover that both first and higher-order topological modes can coexist within the same bulk bandgap. By controlled manipulation of magnetization orientations, we generate Chern insulators with opposing Chern numbers in a bilayer gyromagnetic photonic crystal. Deliberately coupling these distinct layers reveals a higher-order topological phase. This phase creates a bandgap in the initially gapless chiral edge states while allowing the emergence of topological corner modes within this gap. These modes are related to the Jackiw-Rebbi topological bound modes, which appear due to a phase disparity, leading to their localization at specific corners of the sample. Our findings introduce a novel approach to achieving HOTIs by utilizing chiral edge states. This method binds the topological mode within the bulk bandgap and exhibits superior robustness against defects. This advancement holds significant promise for the robust localization of light, presenting opportunities for applications in various fields. Overall, this work enhances the understanding and realization of HOTIs using chiral edge states and offers valuable insights into the robust manipulation of these states for a wide range of applications.

\vspace{15pt}
\noindent \textbf{\large{Methods}}
\vspace{5pt}

\noindent\textbf{Materials and experimental setup} 

Yttrium iron garnet (YIG) ferrites are widely employed as gyromagnetic materials in various commercial applications. These YIG ferrites exhibit a saturation magnetization ($M_s$) of approximately 1780 Gauss, coupled with a gyromagnetic resonance loss width of 35 Oe. Additionally, their permittivity $(\epsilon_r)$ is approximately $14.3+0.0002i$, while their permeability $(\mu_r)$ is approximately $1$, remaining essentially constant at microwave frequencies. The permanent magnets used in this study are composed of Sm$_2$Co$_{17}$ and are electroplated with a nickel layer measuring 0.002 mm in thickness. By utilizing two pairs of magnets, an overall uniform external magnetic field of approximately 0.22 Tesla is generated to induce a gyromagnetic response in the gyromagnetic rods. To ensure precise perforation of the metallic plates, a laser cutting technique is employed, creating air holes. To secure the positions of the sandwiched YIG rods and permanent magnets, dielectric foams known as ROHACELL 31 HF are utilized. These foams have a relative permittivity of 1.04 and a loss tangent of 0.0025. Experimental measurements involve the use of two microwave dipole antennas, serving as the source and probe, respectively. The antennas are connected to a vector network analyzer (Keysight E5080). Finally, we keep the source position fixed and sequentially insert the probe into each of the air holes to obtain the distributions of the electric field.

\vspace{5pt}

\noindent\textbf{Numerical simulation} 

The numerical results in this study were obtained through simulations using the RF module of COMSOL Multiphysics. The strip bandstructures are calculated using $1\times23$ supercell and applying periodic boundary conditions along the x direction and PEC boundary conditions along the y direction. The copper plates and permanent magnets are both considered as PEC. The permeability tensor of the magnetized gyromagnetic material is given by 
\begin{align*}
&\left[\mu_r\right]=
\begin{bmatrix}
\mu_m        & \pm i\mu_k & 0 \\
\mp i\mu_k & \mu_m        & 0 \\
0                & 0                  &1
\end{bmatrix},~
\mu_m=1+\frac{\omega_m(\omega_0+i \alpha \omega)}{[(\omega_0+i \alpha \omega)^2-\omega^2]},\notag\\
&\mu_k=\frac{\omega_m\omega}{[(\omega_0+i \alpha \omega)^2-\omega^2]},~\omega_m=\mu_0\gamma M_s,~\omega_0=\gamma\mu_0 H_0,
\end{align*}
and $\mu_0 H_0$ is the external magnetic field along the $z$ direction. $\gamma=1.759\times10^{11}$ C/kg is the gyromagnetic ratio, $\alpha=0.0088$ is the damping coefficient, and $\omega$ is the operating frequency \cite{Pozar}.

\vspace{5pt}

\noindent\textbf{Data availability}

The data that support the plots within this paper and other findings of this study are available from the corresponding author upon reasonable request.

\bibliography{Maintext}

\begin{thebibliography}{34}%
\makeatletter
\providecommand \@ifxundefined [1]{%
 \@ifx{#1\undefined}
}%
\providecommand \@ifnum [1]{%
 \ifnum #1\expandafter \@firstoftwo
 \else \expandafter \@secondoftwo
 \fi
}%
\providecommand \@ifx [1]{%
 \ifx #1\expandafter \@firstoftwo
 \else \expandafter \@secondoftwo
 \fi
}%
\providecommand \natexlab [1]{#1}%
\providecommand \enquote  [1]{``#1''}%
\providecommand \bibnamefont  [1]{#1}%
\providecommand \bibfnamefont [1]{#1}%
\providecommand \citenamefont [1]{#1}%
\providecommand \href@noop [0]{\@secondoftwo}%
\providecommand \href [0]{\begingroup \@sanitize@url \@href}%
\providecommand \@href[1]{\@@startlink{#1}\@@href}%
\providecommand \@@href[1]{\endgroup#1\@@endlink}%
\providecommand \@sanitize@url [0]{\catcode `\\12\catcode `\$12\catcode
  `\&12\catcode `\#12\catcode `\^12\catcode `\_12\catcode `\%12\relax}%
\providecommand \@@startlink[1]{}%
\providecommand \@@endlink[0]{}%
\providecommand \url  [0]{\begingroup\@sanitize@url \@url }%
\providecommand \@url [1]{\endgroup\@href {#1}{\urlprefix }}%
\providecommand \urlprefix  [0]{URL }%
\providecommand \Eprint [0]{\href }%
\providecommand \doibase [0]{https://doi.org/}%
\providecommand \selectlanguage [0]{\@gobble}%
\providecommand \bibinfo  [0]{\@secondoftwo}%
\providecommand \bibfield  [0]{\@secondoftwo}%
\providecommand \translation [1]{[#1]}%
\providecommand \BibitemOpen [0]{}%
\providecommand \bibitemStop [0]{}%
\providecommand \bibitemNoStop [0]{.\EOS\space}%
\providecommand \EOS [0]{\spacefactor3000\relax}%
\providecommand \BibitemShut  [1]{\csname bibitem#1\endcsname}%
\let\auto@bib@innerbib\@empty
\bibitem [{\citenamefont {Hasan}\ and\ \citenamefont
  {Kane}(2010)}]{RevModPhys.82.3045}%
  \BibitemOpen
  \bibfield  {author} {\bibinfo {author} {\bibfnamefont {M.~Z.}\ \bibnamefont
  {Hasan}}\ and\ \bibinfo {author} {\bibfnamefont {C.~L.}\ \bibnamefont
  {Kane}},\ }\bibfield  {title} {\bibinfo {title} {Colloquium: Topological
  insulators},\ }\href {https://doi.org/10.1103/RevModPhys.82.3045} {\bibfield
  {journal} {\bibinfo  {journal} {Rev. Mod. Phys.}\ }\textbf {\bibinfo {volume}
  {82}},\ \bibinfo {pages} {3045} (\bibinfo {year} {2010})}\BibitemShut
  {NoStop}%
\bibitem [{\citenamefont {Lu}\ \emph {et~al.}(2014)\citenamefont {Lu},
  \citenamefont {Joannopoulos},\ and\ \citenamefont
  {Solja{\v{c}}i{\'{c}}}}]{Lu2014}%
  \BibitemOpen
  \bibfield  {author} {\bibinfo {author} {\bibfnamefont {L.}~\bibnamefont
  {Lu}}, \bibinfo {author} {\bibfnamefont {J.~D.}\ \bibnamefont
  {Joannopoulos}},\ and\ \bibinfo {author} {\bibfnamefont {M.}~\bibnamefont
  {Solja{\v{c}}i{\'{c}}}},\ }\bibfield  {title} {\bibinfo {title} {{Topological
  photonics}},\ }\href {https://doi.org/10.1038/nphoton.2014.248} {\bibfield
  {journal} {\bibinfo  {journal} {Nature Photonics}\ }\textbf {\bibinfo
  {volume} {8}},\ \bibinfo {pages} {821} (\bibinfo {year} {2014})}\BibitemShut
  {NoStop}%
\bibitem [{\citenamefont {Khanikaev}\ and\ \citenamefont
  {Shvets}(2017)}]{Khanikaev2017}%
  \BibitemOpen
  \bibfield  {author} {\bibinfo {author} {\bibfnamefont {A.~B.}\ \bibnamefont
  {Khanikaev}}\ and\ \bibinfo {author} {\bibfnamefont {G.}~\bibnamefont
  {Shvets}},\ }\bibfield  {title} {\bibinfo {title} {{Two-dimensional
  topological photonics}},\ }\href {https://doi.org/10.1038/s41566-017-0048-5}
  {\bibfield  {journal} {\bibinfo  {journal} {Nature Photonics}\ }\textbf
  {\bibinfo {volume} {11}},\ \bibinfo {pages} {763} (\bibinfo {year}
  {2017})}\BibitemShut {NoStop}%
\bibitem [{\citenamefont {Ozawa}\ \emph {et~al.}(2019)\citenamefont {Ozawa},
  \citenamefont {Price}, \citenamefont {Amo}, \citenamefont {Goldman},
  \citenamefont {Hafezi}, \citenamefont {Lu}, \citenamefont {Rechtsman},
  \citenamefont {Schuster}, \citenamefont {Simon}, \citenamefont {Zilberberg},\
  and\ \citenamefont {Carusotto}}]{RevModPhys.91.015006}%
  \BibitemOpen
  \bibfield  {author} {\bibinfo {author} {\bibfnamefont {T.}~\bibnamefont
  {Ozawa}}, \bibinfo {author} {\bibfnamefont {H.~M.}\ \bibnamefont {Price}},
  \bibinfo {author} {\bibfnamefont {A.}~\bibnamefont {Amo}}, \bibinfo {author}
  {\bibfnamefont {N.}~\bibnamefont {Goldman}}, \bibinfo {author} {\bibfnamefont
  {M.}~\bibnamefont {Hafezi}}, \bibinfo {author} {\bibfnamefont
  {L.}~\bibnamefont {Lu}}, \bibinfo {author} {\bibfnamefont {M.~C.}\
  \bibnamefont {Rechtsman}}, \bibinfo {author} {\bibfnamefont {D.}~\bibnamefont
  {Schuster}}, \bibinfo {author} {\bibfnamefont {J.}~\bibnamefont {Simon}},
  \bibinfo {author} {\bibfnamefont {O.}~\bibnamefont {Zilberberg}},\ and\
  \bibinfo {author} {\bibfnamefont {I.}~\bibnamefont {Carusotto}},\ }\bibfield
  {title} {\bibinfo {title} {Topological photonics},\ }\href
  {https://doi.org/10.1103/RevModPhys.91.015006} {\bibfield  {journal}
  {\bibinfo  {journal} {Rev. Mod. Phys.}\ }\textbf {\bibinfo {volume} {91}},\
  \bibinfo {pages} {015006} (\bibinfo {year} {2019})}\BibitemShut {NoStop}%
\bibitem [{\citenamefont {Zhan}\ \emph {et~al.}(2023)\citenamefont {Zhan},
  \citenamefont {Qin}, \citenamefont {Xu}, \citenamefont {Zhou}, \citenamefont
  {Ma},\ and\ \citenamefont {Wang}}]{zhan2023design}%
  \BibitemOpen
  \bibfield  {author} {\bibinfo {author} {\bibfnamefont {F.}~\bibnamefont
  {Zhan}}, \bibinfo {author} {\bibfnamefont {Z.}~\bibnamefont {Qin}}, \bibinfo
  {author} {\bibfnamefont {D.-H.}\ \bibnamefont {Xu}}, \bibinfo {author}
  {\bibfnamefont {X.}~\bibnamefont {Zhou}}, \bibinfo {author} {\bibfnamefont
  {D.-S.}\ \bibnamefont {Ma}},\ and\ \bibinfo {author} {\bibfnamefont
  {R.}~\bibnamefont {Wang}},\ }\href@noop {} {\bibinfo {title} {Design of
  antiferromagnetic second-order band topology with rotation topological
  invariants in two dimensions}} (\bibinfo {year} {2023}),\ \Eprint
  {https://arxiv.org/abs/2307.06903} {arXiv:2307.06903 [cond-mat.mtrl-sci]}
  \BibitemShut {NoStop}%
\bibitem [{\citenamefont {Liu}\ \emph {et~al.}(2024{\natexlab{a}})\citenamefont
  {Liu}, \citenamefont {An}, \citenamefont {Ren}, \citenamefont {Zhang},
  \citenamefont {Qiao},\ and\ \citenamefont {Niu}}]{liu2024interlayer}%
  \BibitemOpen
  \bibfield  {author} {\bibinfo {author} {\bibfnamefont {L.}~\bibnamefont
  {Liu}}, \bibinfo {author} {\bibfnamefont {J.}~\bibnamefont {An}}, \bibinfo
  {author} {\bibfnamefont {Y.}~\bibnamefont {Ren}}, \bibinfo {author}
  {\bibfnamefont {Y.}~\bibnamefont {Zhang}}, \bibinfo {author} {\bibfnamefont
  {Z.}~\bibnamefont {Qiao}},\ and\ \bibinfo {author} {\bibfnamefont
  {Q.}~\bibnamefont {Niu}},\ }\href@noop {} {\bibinfo {title} {Interlayer
  coupling induced topological phase transition to higher order}} (\bibinfo
  {year} {2024}{\natexlab{a}}),\ \Eprint {https://arxiv.org/abs/2405.11249}
  {arXiv:2405.11249 [cond-mat.mes-hall]} \BibitemShut {NoStop}%
\bibitem [{\citenamefont {Benalcazar}\ \emph
  {et~al.}(2017{\natexlab{a}})\citenamefont {Benalcazar}, \citenamefont
  {Bernevig},\ and\ \citenamefont {Hughes}}]{doi:10.1126/science.aah6442}%
  \BibitemOpen
  \bibfield  {author} {\bibinfo {author} {\bibfnamefont {W.~A.}\ \bibnamefont
  {Benalcazar}}, \bibinfo {author} {\bibfnamefont {B.~A.}\ \bibnamefont
  {Bernevig}},\ and\ \bibinfo {author} {\bibfnamefont {T.~L.}\ \bibnamefont
  {Hughes}},\ }\bibfield  {title} {\bibinfo {title} {Quantized electric
  multipole insulators},\ }\href {https://doi.org/10.1126/science.aah6442}
  {\bibfield  {journal} {\bibinfo  {journal} {Science}\ }\textbf {\bibinfo
  {volume} {357}},\ \bibinfo {pages} {61} (\bibinfo {year}
  {2017}{\natexlab{a}})}\BibitemShut {NoStop}%
\bibitem [{\citenamefont {Benalcazar}\ \emph
  {et~al.}(2017{\natexlab{b}})\citenamefont {Benalcazar}, \citenamefont
  {Bernevig},\ and\ \citenamefont {Hughes}}]{PhysRevB.96.245115}%
  \BibitemOpen
  \bibfield  {author} {\bibinfo {author} {\bibfnamefont {W.~A.}\ \bibnamefont
  {Benalcazar}}, \bibinfo {author} {\bibfnamefont {B.~A.}\ \bibnamefont
  {Bernevig}},\ and\ \bibinfo {author} {\bibfnamefont {T.~L.}\ \bibnamefont
  {Hughes}},\ }\bibfield  {title} {\bibinfo {title} {Electric multipole
  moments, topological multipole moment pumping, and chiral hinge states in
  crystalline insulators},\ }\href {https://doi.org/10.1103/PhysRevB.96.245115}
  {\bibfield  {journal} {\bibinfo  {journal} {Phys. Rev. B}\ }\textbf {\bibinfo
  {volume} {96}},\ \bibinfo {pages} {245115} (\bibinfo {year}
  {2017}{\natexlab{b}})}\BibitemShut {NoStop}%
\bibitem [{\citenamefont {Xie}\ \emph {et~al.}(2021)\citenamefont {Xie},
  \citenamefont {Wang}, \citenamefont {Zhang}, \citenamefont {Zhan},
  \citenamefont {Jiang}, \citenamefont {Lu},\ and\ \citenamefont
  {Chen}}]{Xie2021}%
  \BibitemOpen
  \bibfield  {author} {\bibinfo {author} {\bibfnamefont {B.}~\bibnamefont
  {Xie}}, \bibinfo {author} {\bibfnamefont {H.-X.}\ \bibnamefont {Wang}},
  \bibinfo {author} {\bibfnamefont {X.}~\bibnamefont {Zhang}}, \bibinfo
  {author} {\bibfnamefont {P.}~\bibnamefont {Zhan}}, \bibinfo {author}
  {\bibfnamefont {J.-H.}\ \bibnamefont {Jiang}}, \bibinfo {author}
  {\bibfnamefont {M.}~\bibnamefont {Lu}},\ and\ \bibinfo {author}
  {\bibfnamefont {Y.}~\bibnamefont {Chen}},\ }\bibfield  {title} {\bibinfo
  {title} {{Higher-order band topology}},\ }\href
  {https://doi.org/10.1038/s42254-021-00323-4} {\bibfield  {journal} {\bibinfo
  {journal} {Nature Reviews Physics}\ }\textbf {\bibinfo {volume} {3}},\
  \bibinfo {pages} {520} (\bibinfo {year} {2021})}\BibitemShut {NoStop}%
\bibitem [{\citenamefont {Peterson}\ \emph {et~al.}(2018)\citenamefont
  {Peterson}, \citenamefont {Benalcazar}, \citenamefont {Hughes},\ and\
  \citenamefont {Bahl}}]{Peterson2018}%
  \BibitemOpen
  \bibfield  {author} {\bibinfo {author} {\bibfnamefont {C.~W.}\ \bibnamefont
  {Peterson}}, \bibinfo {author} {\bibfnamefont {W.~A.}\ \bibnamefont
  {Benalcazar}}, \bibinfo {author} {\bibfnamefont {T.~L.}\ \bibnamefont
  {Hughes}},\ and\ \bibinfo {author} {\bibfnamefont {G.}~\bibnamefont {Bahl}},\
  }\bibfield  {title} {\bibinfo {title} {{A quantized microwave quadrupole
  insulator with topologically protected corner states}},\ }\href
  {https://doi.org/10.1038/nature25777} {\bibfield  {journal} {\bibinfo
  {journal} {Nature}\ }\textbf {\bibinfo {volume} {555}},\ \bibinfo {pages}
  {346} (\bibinfo {year} {2018})}\BibitemShut {NoStop}%
\bibitem [{\citenamefont {Noh}\ \emph {et~al.}(2018)\citenamefont {Noh},
  \citenamefont {Benalcazar}, \citenamefont {Huang}, \citenamefont {Collins},
  \citenamefont {Chen}, \citenamefont {Hughes},\ and\ \citenamefont
  {Rechtsman}}]{Noh2018}%
  \BibitemOpen
  \bibfield  {author} {\bibinfo {author} {\bibfnamefont {J.}~\bibnamefont
  {Noh}}, \bibinfo {author} {\bibfnamefont {W.~A.}\ \bibnamefont {Benalcazar}},
  \bibinfo {author} {\bibfnamefont {S.}~\bibnamefont {Huang}}, \bibinfo
  {author} {\bibfnamefont {M.~J.}\ \bibnamefont {Collins}}, \bibinfo {author}
  {\bibfnamefont {K.~P.}\ \bibnamefont {Chen}}, \bibinfo {author}
  {\bibfnamefont {T.~L.}\ \bibnamefont {Hughes}},\ and\ \bibinfo {author}
  {\bibfnamefont {M.~C.}\ \bibnamefont {Rechtsman}},\ }\bibfield  {title}
  {\bibinfo {title} {{Topological protection of photonic mid-gap defect
  modes}},\ }\href {https://doi.org/10.1038/s41566-018-0179-3} {\bibfield
  {journal} {\bibinfo  {journal} {Nature Photonics}\ }\textbf {\bibinfo
  {volume} {12}},\ \bibinfo {pages} {408} (\bibinfo {year} {2018})}\BibitemShut
  {NoStop}%
\bibitem [{\citenamefont {Mittal}\ \emph {et~al.}(2019)\citenamefont {Mittal},
  \citenamefont {Orre}, \citenamefont {Zhu}, \citenamefont {Gorlach},
  \citenamefont {Poddubny},\ and\ \citenamefont {Hafezi}}]{Mittal2019}%
  \BibitemOpen
  \bibfield  {author} {\bibinfo {author} {\bibfnamefont {S.}~\bibnamefont
  {Mittal}}, \bibinfo {author} {\bibfnamefont {V.~V.}\ \bibnamefont {Orre}},
  \bibinfo {author} {\bibfnamefont {G.}~\bibnamefont {Zhu}}, \bibinfo {author}
  {\bibfnamefont {M.~A.}\ \bibnamefont {Gorlach}}, \bibinfo {author}
  {\bibfnamefont {A.}~\bibnamefont {Poddubny}},\ and\ \bibinfo {author}
  {\bibfnamefont {M.}~\bibnamefont {Hafezi}},\ }\bibfield  {title} {\bibinfo
  {title} {{Photonic quadrupole topological phases}},\ }\href
  {https://doi.org/10.1038/s41566-019-0452-0} {\bibfield  {journal} {\bibinfo
  {journal} {Nature Photonics}\ }\textbf {\bibinfo {volume} {13}},\ \bibinfo
  {pages} {692} (\bibinfo {year} {2019})}\BibitemShut {NoStop}%
\bibitem [{\citenamefont {Cerjan}\ \emph {et~al.}(2020)\citenamefont {Cerjan},
  \citenamefont {J\"urgensen}, \citenamefont {Benalcazar}, \citenamefont
  {Mukherjee},\ and\ \citenamefont {Rechtsman}}]{PhysRevLett.125.213901}%
  \BibitemOpen
  \bibfield  {author} {\bibinfo {author} {\bibfnamefont {A.}~\bibnamefont
  {Cerjan}}, \bibinfo {author} {\bibfnamefont {M.}~\bibnamefont {J\"urgensen}},
  \bibinfo {author} {\bibfnamefont {W.~A.}\ \bibnamefont {Benalcazar}},
  \bibinfo {author} {\bibfnamefont {S.}~\bibnamefont {Mukherjee}},\ and\
  \bibinfo {author} {\bibfnamefont {M.~C.}\ \bibnamefont {Rechtsman}},\
  }\bibfield  {title} {\bibinfo {title} {Observation of a higher-order
  topological bound state in the continuum},\ }\href
  {https://doi.org/10.1103/PhysRevLett.125.213901} {\bibfield  {journal}
  {\bibinfo  {journal} {Phys. Rev. Lett.}\ }\textbf {\bibinfo {volume} {125}},\
  \bibinfo {pages} {213901} (\bibinfo {year} {2020})}\BibitemShut {NoStop}%
\bibitem [{\citenamefont {Xie}\ \emph {et~al.}(2020)\citenamefont {Xie},
  \citenamefont {Su}, \citenamefont {Wang}, \citenamefont {Liu}, \citenamefont
  {Hu}, \citenamefont {Yu}, \citenamefont {Zhan}, \citenamefont {Lu},
  \citenamefont {Wang},\ and\ \citenamefont {Chen}}]{Xie2020}%
  \BibitemOpen
  \bibfield  {author} {\bibinfo {author} {\bibfnamefont {B.}~\bibnamefont
  {Xie}}, \bibinfo {author} {\bibfnamefont {G.}~\bibnamefont {Su}}, \bibinfo
  {author} {\bibfnamefont {H.-F.}\ \bibnamefont {Wang}}, \bibinfo {author}
  {\bibfnamefont {F.}~\bibnamefont {Liu}}, \bibinfo {author} {\bibfnamefont
  {L.}~\bibnamefont {Hu}}, \bibinfo {author} {\bibfnamefont {S.-Y.}\
  \bibnamefont {Yu}}, \bibinfo {author} {\bibfnamefont {P.}~\bibnamefont
  {Zhan}}, \bibinfo {author} {\bibfnamefont {M.-H.}\ \bibnamefont {Lu}},
  \bibinfo {author} {\bibfnamefont {Z.}~\bibnamefont {Wang}},\ and\ \bibinfo
  {author} {\bibfnamefont {Y.-F.}\ \bibnamefont {Chen}},\ }\bibfield  {title}
  {\bibinfo {title} {{Higher-order quantum spin Hall effect in a photonic
  crystal}},\ }\href {https://doi.org/10.1038/s41467-020-17593-8} {\bibfield
  {journal} {\bibinfo  {journal} {Nature Communications}\ }\textbf {\bibinfo
  {volume} {11}},\ \bibinfo {pages} {3768} (\bibinfo {year}
  {2020})}\BibitemShut {NoStop}%
\bibitem [{\citenamefont {Schulz}\ \emph {et~al.}(2022)\citenamefont {Schulz},
  \citenamefont {Noh}, \citenamefont {Benalcazar}, \citenamefont {Bahl},\ and\
  \citenamefont {von Freymann}}]{Schulz2022}%
  \BibitemOpen
  \bibfield  {author} {\bibinfo {author} {\bibfnamefont {J.}~\bibnamefont
  {Schulz}}, \bibinfo {author} {\bibfnamefont {J.}~\bibnamefont {Noh}},
  \bibinfo {author} {\bibfnamefont {W.~A.}\ \bibnamefont {Benalcazar}},
  \bibinfo {author} {\bibfnamefont {G.}~\bibnamefont {Bahl}},\ and\ \bibinfo
  {author} {\bibfnamefont {G.}~\bibnamefont {von Freymann}},\ }\bibfield
  {title} {\bibinfo {title} {{Photonic quadrupole topological insulator using
  orbital-induced synthetic flux}},\ }\href
  {https://doi.org/10.1038/s41467-022-33894-6} {\bibfield  {journal} {\bibinfo
  {journal} {Nature Communications}\ }\textbf {\bibinfo {volume} {13}},\
  \bibinfo {pages} {6597} (\bibinfo {year} {2022})}\BibitemShut {NoStop}%
\bibitem [{\citenamefont {Ota}\ \emph {et~al.}(2019)\citenamefont {Ota},
  \citenamefont {Liu}, \citenamefont {Katsumi}, \citenamefont {Watanabe},
  \citenamefont {Wakabayashi}, \citenamefont {Arakawa},\ and\ \citenamefont
  {Iwamoto}}]{Ota:19}%
  \BibitemOpen
  \bibfield  {author} {\bibinfo {author} {\bibfnamefont {Y.}~\bibnamefont
  {Ota}}, \bibinfo {author} {\bibfnamefont {F.}~\bibnamefont {Liu}}, \bibinfo
  {author} {\bibfnamefont {R.}~\bibnamefont {Katsumi}}, \bibinfo {author}
  {\bibfnamefont {K.}~\bibnamefont {Watanabe}}, \bibinfo {author}
  {\bibfnamefont {K.}~\bibnamefont {Wakabayashi}}, \bibinfo {author}
  {\bibfnamefont {Y.}~\bibnamefont {Arakawa}},\ and\ \bibinfo {author}
  {\bibfnamefont {S.}~\bibnamefont {Iwamoto}},\ }\bibfield  {title} {\bibinfo
  {title} {Photonic crystal nanocavity based on a topological corner state},\
  }\href {https://doi.org/10.1364/OPTICA.6.000786} {\bibfield  {journal}
  {\bibinfo  {journal} {Optica}\ }\textbf {\bibinfo {volume} {6}},\ \bibinfo
  {pages} {786} (\bibinfo {year} {2019})}\BibitemShut {NoStop}%
\bibitem [{\citenamefont {Zhang}\ \emph {et~al.}(2020)\citenamefont {Zhang},
  \citenamefont {Xie}, \citenamefont {Hao}, \citenamefont {Dang}, \citenamefont
  {Xiao}, \citenamefont {Shi}, \citenamefont {Ni}, \citenamefont {Niu},
  \citenamefont {Wang}, \citenamefont {Jin}, \citenamefont {Zhang},\ and\
  \citenamefont {Xu}}]{Zhang2020}%
  \BibitemOpen
  \bibfield  {author} {\bibinfo {author} {\bibfnamefont {W.}~\bibnamefont
  {Zhang}}, \bibinfo {author} {\bibfnamefont {X.}~\bibnamefont {Xie}}, \bibinfo
  {author} {\bibfnamefont {H.}~\bibnamefont {Hao}}, \bibinfo {author}
  {\bibfnamefont {J.}~\bibnamefont {Dang}}, \bibinfo {author} {\bibfnamefont
  {S.}~\bibnamefont {Xiao}}, \bibinfo {author} {\bibfnamefont {S.}~\bibnamefont
  {Shi}}, \bibinfo {author} {\bibfnamefont {H.}~\bibnamefont {Ni}}, \bibinfo
  {author} {\bibfnamefont {Z.}~\bibnamefont {Niu}}, \bibinfo {author}
  {\bibfnamefont {C.}~\bibnamefont {Wang}}, \bibinfo {author} {\bibfnamefont
  {K.}~\bibnamefont {Jin}}, \bibinfo {author} {\bibfnamefont {X.}~\bibnamefont
  {Zhang}},\ and\ \bibinfo {author} {\bibfnamefont {X.}~\bibnamefont {Xu}},\
  }\bibfield  {title} {\bibinfo {title} {Low-threshold topological nanolasers
  based on the second-order corner state},\ }\href
  {https://doi.org/10.1038/s41377-020-00352-1} {\bibfield  {journal} {\bibinfo
  {journal} {Light: Science \& Applications}\ }\textbf {\bibinfo {volume}
  {9}},\ \bibinfo {pages} {109} (\bibinfo {year} {2020})}\BibitemShut {NoStop}%
\bibitem [{\citenamefont {Kim}\ \emph {et~al.}(2020)\citenamefont {Kim},
  \citenamefont {Hwang}, \citenamefont {Smirnova}, \citenamefont {Jeong},
  \citenamefont {Kivshar},\ and\ \citenamefont {Park}}]{Kim2020}%
  \BibitemOpen
  \bibfield  {author} {\bibinfo {author} {\bibfnamefont {H.-R.}\ \bibnamefont
  {Kim}}, \bibinfo {author} {\bibfnamefont {M.-S.}\ \bibnamefont {Hwang}},
  \bibinfo {author} {\bibfnamefont {D.}~\bibnamefont {Smirnova}}, \bibinfo
  {author} {\bibfnamefont {K.-Y.}\ \bibnamefont {Jeong}}, \bibinfo {author}
  {\bibfnamefont {Y.}~\bibnamefont {Kivshar}},\ and\ \bibinfo {author}
  {\bibfnamefont {H.-G.}\ \bibnamefont {Park}},\ }\bibfield  {title} {\bibinfo
  {title} {{Multipolar lasing modes from topological corner states}},\ }\href
  {https://doi.org/10.1038/s41467-020-19609-9} {\bibfield  {journal} {\bibinfo
  {journal} {Nature Communications}\ }\textbf {\bibinfo {volume} {11}},\
  \bibinfo {pages} {5758} (\bibinfo {year} {2020})}\BibitemShut {NoStop}%
\bibitem [{\citenamefont {Wu}\ \emph {et~al.}(2023)\citenamefont {Wu},
  \citenamefont {Ghosh}, \citenamefont {Gan}, \citenamefont {Shi},
  \citenamefont {Mandal}, \citenamefont {Sun}, \citenamefont {Zhang},
  \citenamefont {Liew}, \citenamefont {Su},\ and\ \citenamefont
  {Xiong}}]{doi:10.1126/sciadv.adg4322}%
  \BibitemOpen
  \bibfield  {author} {\bibinfo {author} {\bibfnamefont {J.}~\bibnamefont
  {Wu}}, \bibinfo {author} {\bibfnamefont {S.}~\bibnamefont {Ghosh}}, \bibinfo
  {author} {\bibfnamefont {Y.}~\bibnamefont {Gan}}, \bibinfo {author}
  {\bibfnamefont {Y.}~\bibnamefont {Shi}}, \bibinfo {author} {\bibfnamefont
  {S.}~\bibnamefont {Mandal}}, \bibinfo {author} {\bibfnamefont
  {H.}~\bibnamefont {Sun}}, \bibinfo {author} {\bibfnamefont {B.}~\bibnamefont
  {Zhang}}, \bibinfo {author} {\bibfnamefont {T.~C.~H.}\ \bibnamefont {Liew}},
  \bibinfo {author} {\bibfnamefont {R.}~\bibnamefont {Su}},\ and\ \bibinfo
  {author} {\bibfnamefont {Q.}~\bibnamefont {Xiong}},\ }\bibfield  {title}
  {\bibinfo {title} {Higher-order topological polariton corner state lasing},\
  }\href {https://doi.org/10.1126/sciadv.adg4322} {\bibfield  {journal}
  {\bibinfo  {journal} {Science Advances}\ }\textbf {\bibinfo {volume} {9}},\
  \bibinfo {pages} {eadg4322} (\bibinfo {year} {2023})}\BibitemShut {NoStop}%
\bibitem [{\citenamefont {Bennenhei}\ \emph {et~al.}(2024)\citenamefont
  {Bennenhei}, \citenamefont {Shan}, \citenamefont {Struve}, \citenamefont
  {Kunte}, \citenamefont {Eilenberger}, \citenamefont {Ohmer}, \citenamefont
  {Fischer}, \citenamefont {Schumacher}, \citenamefont {Ma}, \citenamefont
  {Schneider},\ and\ \citenamefont {Esmann}}]{bennenhei2024organic}%
  \BibitemOpen
  \bibfield  {author} {\bibinfo {author} {\bibfnamefont {C.}~\bibnamefont
  {Bennenhei}}, \bibinfo {author} {\bibfnamefont {H.}~\bibnamefont {Shan}},
  \bibinfo {author} {\bibfnamefont {M.}~\bibnamefont {Struve}}, \bibinfo
  {author} {\bibfnamefont {N.}~\bibnamefont {Kunte}}, \bibinfo {author}
  {\bibfnamefont {F.}~\bibnamefont {Eilenberger}}, \bibinfo {author}
  {\bibfnamefont {J.}~\bibnamefont {Ohmer}}, \bibinfo {author} {\bibfnamefont
  {U.}~\bibnamefont {Fischer}}, \bibinfo {author} {\bibfnamefont
  {S.}~\bibnamefont {Schumacher}}, \bibinfo {author} {\bibfnamefont
  {X.}~\bibnamefont {Ma}}, \bibinfo {author} {\bibfnamefont {C.}~\bibnamefont
  {Schneider}},\ and\ \bibinfo {author} {\bibfnamefont {M.}~\bibnamefont
  {Esmann}},\ }\href@noop {} {\bibinfo {title} {Organic room-temperature
  polariton condensate in a higher-order topological lattice}} (\bibinfo {year}
  {2024}),\ \Eprint {https://arxiv.org/abs/2401.06267} {arXiv:2401.06267
  [physics.optics]} \BibitemShut {NoStop}%
\bibitem [{\citenamefont {Peterson}\ \emph {et~al.}(2020)\citenamefont
  {Peterson}, \citenamefont {Li}, \citenamefont {Benalcazar}, \citenamefont
  {Hughes},\ and\ \citenamefont {Bahl}}]{doi:10.1126/science.aba7604}%
  \BibitemOpen
  \bibfield  {author} {\bibinfo {author} {\bibfnamefont {C.~W.}\ \bibnamefont
  {Peterson}}, \bibinfo {author} {\bibfnamefont {T.}~\bibnamefont {Li}},
  \bibinfo {author} {\bibfnamefont {W.~A.}\ \bibnamefont {Benalcazar}},
  \bibinfo {author} {\bibfnamefont {T.~L.}\ \bibnamefont {Hughes}},\ and\
  \bibinfo {author} {\bibfnamefont {G.}~\bibnamefont {Bahl}},\ }\bibfield
  {title} {\bibinfo {title} {A fractional corner anomaly reveals higher-order
  topology},\ }\href {https://doi.org/10.1126/science.aba7604} {\bibfield
  {journal} {\bibinfo  {journal} {Science}\ }\textbf {\bibinfo {volume}
  {368}},\ \bibinfo {pages} {1114} (\bibinfo {year} {2020})}\BibitemShut
  {NoStop}%
\bibitem [{\citenamefont {Wang}\ \emph {et~al.}(2009)\citenamefont {Wang},
  \citenamefont {Chong}, \citenamefont {Joannopoulos},\ and\ \citenamefont
  {Solja{\v{c}}i{\'{c}}}}]{Wang2009}%
  \BibitemOpen
  \bibfield  {author} {\bibinfo {author} {\bibfnamefont {Z.}~\bibnamefont
  {Wang}}, \bibinfo {author} {\bibfnamefont {Y.}~\bibnamefont {Chong}},
  \bibinfo {author} {\bibfnamefont {J.~D.}\ \bibnamefont {Joannopoulos}},\ and\
  \bibinfo {author} {\bibfnamefont {M.}~\bibnamefont {Solja{\v{c}}i{\'{c}}}},\
  }\bibfield  {title} {\bibinfo {title} {{Observation of unidirectional
  backscattering-immune topological electromagnetic states}},\ }\href
  {https://doi.org/10.1038/nature08293} {\bibfield  {journal} {\bibinfo
  {journal} {Nature}\ }\textbf {\bibinfo {volume} {461}},\ \bibinfo {pages}
  {772} (\bibinfo {year} {2009})}\BibitemShut {NoStop}%
\bibitem [{\citenamefont {Poo}\ \emph {et~al.}(2011)\citenamefont {Poo},
  \citenamefont {Wu}, \citenamefont {Lin}, \citenamefont {Yang},\ and\
  \citenamefont {Chan}}]{PhysRevLett.106.093903}%
  \BibitemOpen
  \bibfield  {author} {\bibinfo {author} {\bibfnamefont {Y.}~\bibnamefont
  {Poo}}, \bibinfo {author} {\bibfnamefont {R.-x.}\ \bibnamefont {Wu}},
  \bibinfo {author} {\bibfnamefont {Z.}~\bibnamefont {Lin}}, \bibinfo {author}
  {\bibfnamefont {Y.}~\bibnamefont {Yang}},\ and\ \bibinfo {author}
  {\bibfnamefont {C.~T.}\ \bibnamefont {Chan}},\ }\bibfield  {title} {\bibinfo
  {title} {Experimental realization of self-guiding unidirectional
  electromagnetic edge states},\ }\href
  {https://doi.org/10.1103/PhysRevLett.106.093903} {\bibfield  {journal}
  {\bibinfo  {journal} {Phys. Rev. Lett.}\ }\textbf {\bibinfo {volume} {106}},\
  \bibinfo {pages} {093903} (\bibinfo {year} {2011})}\BibitemShut {NoStop}%
\bibitem [{\citenamefont {Liu}\ \emph {et~al.}(2024{\natexlab{b}})\citenamefont
  {Liu}, \citenamefont {Mandal}, \citenamefont {Zhou}, \citenamefont {Xi},
  \citenamefont {Banerjee}, \citenamefont {Hu}, \citenamefont {Wei},
  \citenamefont {Wang}, \citenamefont {Wang}, \citenamefont {Gao},
  \citenamefont {Chen}, \citenamefont {Yang}, \citenamefont {Chong},\ and\
  \citenamefont {Zhang}}]{PhysRevLett.132.113802}%
  \BibitemOpen
  \bibfield  {author} {\bibinfo {author} {\bibfnamefont {G.-G.}\ \bibnamefont
  {Liu}}, \bibinfo {author} {\bibfnamefont {S.}~\bibnamefont {Mandal}},
  \bibinfo {author} {\bibfnamefont {P.}~\bibnamefont {Zhou}}, \bibinfo {author}
  {\bibfnamefont {X.}~\bibnamefont {Xi}}, \bibinfo {author} {\bibfnamefont
  {R.}~\bibnamefont {Banerjee}}, \bibinfo {author} {\bibfnamefont {Y.-H.}\
  \bibnamefont {Hu}}, \bibinfo {author} {\bibfnamefont {M.}~\bibnamefont
  {Wei}}, \bibinfo {author} {\bibfnamefont {M.}~\bibnamefont {Wang}}, \bibinfo
  {author} {\bibfnamefont {Q.}~\bibnamefont {Wang}}, \bibinfo {author}
  {\bibfnamefont {Z.}~\bibnamefont {Gao}}, \bibinfo {author} {\bibfnamefont
  {H.}~\bibnamefont {Chen}}, \bibinfo {author} {\bibfnamefont {Y.}~\bibnamefont
  {Yang}}, \bibinfo {author} {\bibfnamefont {Y.}~\bibnamefont {Chong}},\ and\
  \bibinfo {author} {\bibfnamefont {B.}~\bibnamefont {Zhang}},\ }\bibfield
  {title} {\bibinfo {title} {Localization of chiral edge states by the
  non-hermitian skin effect},\ }\href
  {https://doi.org/10.1103/PhysRevLett.132.113802} {\bibfield  {journal}
  {\bibinfo  {journal} {Phys. Rev. Lett.}\ }\textbf {\bibinfo {volume} {132}},\
  \bibinfo {pages} {113802} (\bibinfo {year} {2024}{\natexlab{b}})}\BibitemShut
  {NoStop}%
\bibitem [{\citenamefont {Zhou}\ \emph {et~al.}(2020)\citenamefont {Zhou},
  \citenamefont {Liu}, \citenamefont {Yang}, \citenamefont {Hu}, \citenamefont
  {Ma}, \citenamefont {Xue}, \citenamefont {Wang}, \citenamefont {Deng},\ and\
  \citenamefont {Zhang}}]{PhysRevLett.125.263603}%
  \BibitemOpen
  \bibfield  {author} {\bibinfo {author} {\bibfnamefont {P.}~\bibnamefont
  {Zhou}}, \bibinfo {author} {\bibfnamefont {G.-G.}\ \bibnamefont {Liu}},
  \bibinfo {author} {\bibfnamefont {Y.}~\bibnamefont {Yang}}, \bibinfo {author}
  {\bibfnamefont {Y.-H.}\ \bibnamefont {Hu}}, \bibinfo {author} {\bibfnamefont
  {S.}~\bibnamefont {Ma}}, \bibinfo {author} {\bibfnamefont {H.}~\bibnamefont
  {Xue}}, \bibinfo {author} {\bibfnamefont {Q.}~\bibnamefont {Wang}}, \bibinfo
  {author} {\bibfnamefont {L.}~\bibnamefont {Deng}},\ and\ \bibinfo {author}
  {\bibfnamefont {B.}~\bibnamefont {Zhang}},\ }\bibfield  {title} {\bibinfo
  {title} {Observation of photonic antichiral edge states},\ }\href
  {https://doi.org/10.1103/PhysRevLett.125.263603} {\bibfield  {journal}
  {\bibinfo  {journal} {Phys. Rev. Lett.}\ }\textbf {\bibinfo {volume} {125}},\
  \bibinfo {pages} {263603} (\bibinfo {year} {2020})}\BibitemShut {NoStop}%
\bibitem [{\citenamefont {Chen}\ and\ \citenamefont
  {Li}(2022)}]{Jianfeng_Chen_2022_Opto_Electronic_Science}%
  \BibitemOpen
  \bibfield  {author} {\bibinfo {author} {\bibfnamefont {J.}~\bibnamefont
  {Chen}}\ and\ \bibinfo {author} {\bibfnamefont {Z.-Y.}\ \bibnamefont {Li}},\
  }\bibfield  {title} {\bibinfo {title} {Configurable topological beam
  splitting via antichiral gyromagnetic photonic crystal},\ }\href
  {https://doi.org/10.29026/oes.2022.220001} {\bibfield  {journal} {\bibinfo
  {journal} {Opto-Electronic Science}\ }\textbf {\bibinfo {volume} {1}},\
  \bibinfo {pages} {220001} (\bibinfo {year} {2022})}\BibitemShut {NoStop}%
\bibitem [{\citenamefont {Xi}\ \emph {et~al.}(2023)\citenamefont {Xi},
  \citenamefont {Yan}, \citenamefont {Yang}, \citenamefont {Meng},
  \citenamefont {Zhu}, \citenamefont {Chen}, \citenamefont {Wang},
  \citenamefont {Zhou}, \citenamefont {Shum}, \citenamefont {Yang},
  \citenamefont {Chen}, \citenamefont {Mandal}, \citenamefont {Liu},
  \citenamefont {Zhang},\ and\ \citenamefont {Gao}}]{Xi2023}%
  \BibitemOpen
  \bibfield  {author} {\bibinfo {author} {\bibfnamefont {X.}~\bibnamefont
  {Xi}}, \bibinfo {author} {\bibfnamefont {B.}~\bibnamefont {Yan}}, \bibinfo
  {author} {\bibfnamefont {L.}~\bibnamefont {Yang}}, \bibinfo {author}
  {\bibfnamefont {Y.}~\bibnamefont {Meng}}, \bibinfo {author} {\bibfnamefont
  {Z.-X.}\ \bibnamefont {Zhu}}, \bibinfo {author} {\bibfnamefont {J.-M.}\
  \bibnamefont {Chen}}, \bibinfo {author} {\bibfnamefont {Z.}~\bibnamefont
  {Wang}}, \bibinfo {author} {\bibfnamefont {P.}~\bibnamefont {Zhou}}, \bibinfo
  {author} {\bibfnamefont {P.~P.}\ \bibnamefont {Shum}}, \bibinfo {author}
  {\bibfnamefont {Y.}~\bibnamefont {Yang}}, \bibinfo {author} {\bibfnamefont
  {H.}~\bibnamefont {Chen}}, \bibinfo {author} {\bibfnamefont {S.}~\bibnamefont
  {Mandal}}, \bibinfo {author} {\bibfnamefont {G.-G.}\ \bibnamefont {Liu}},
  \bibinfo {author} {\bibfnamefont {B.}~\bibnamefont {Zhang}},\ and\ \bibinfo
  {author} {\bibfnamefont {Z.}~\bibnamefont {Gao}},\ }\bibfield  {title}
  {\bibinfo {title} {{Topological antichiral surface states in a magnetic Weyl
  photonic crystal}},\ }\href {https://doi.org/10.1038/s41467-023-37710-7}
  {\bibfield  {journal} {\bibinfo  {journal} {Nature Communications}\ }\textbf
  {\bibinfo {volume} {14}},\ \bibinfo {pages} {1991} (\bibinfo {year}
  {2023})}\BibitemShut {NoStop}%
\bibitem [{\citenamefont {Liu}\ \emph {et~al.}(2020)\citenamefont {Liu},
  \citenamefont {Yang}, \citenamefont {Ren}, \citenamefont {Xue}, \citenamefont
  {Lin}, \citenamefont {Hu}, \citenamefont {Sun}, \citenamefont {Peng},
  \citenamefont {Zhou}, \citenamefont {Chong},\ and\ \citenamefont
  {Zhang}}]{PhysRevLett.125.133603}%
  \BibitemOpen
  \bibfield  {author} {\bibinfo {author} {\bibfnamefont {G.-G.}\ \bibnamefont
  {Liu}}, \bibinfo {author} {\bibfnamefont {Y.}~\bibnamefont {Yang}}, \bibinfo
  {author} {\bibfnamefont {X.}~\bibnamefont {Ren}}, \bibinfo {author}
  {\bibfnamefont {H.}~\bibnamefont {Xue}}, \bibinfo {author} {\bibfnamefont
  {X.}~\bibnamefont {Lin}}, \bibinfo {author} {\bibfnamefont {Y.-H.}\
  \bibnamefont {Hu}}, \bibinfo {author} {\bibfnamefont {H.-x.}\ \bibnamefont
  {Sun}}, \bibinfo {author} {\bibfnamefont {B.}~\bibnamefont {Peng}}, \bibinfo
  {author} {\bibfnamefont {P.}~\bibnamefont {Zhou}}, \bibinfo {author}
  {\bibfnamefont {Y.}~\bibnamefont {Chong}},\ and\ \bibinfo {author}
  {\bibfnamefont {B.}~\bibnamefont {Zhang}},\ }\bibfield  {title} {\bibinfo
  {title} {Topological anderson insulator in disordered photonic crystals},\
  }\href {https://doi.org/10.1103/PhysRevLett.125.133603} {\bibfield  {journal}
  {\bibinfo  {journal} {Phys. Rev. Lett.}\ }\textbf {\bibinfo {volume} {125}},\
  \bibinfo {pages} {133603} (\bibinfo {year} {2020})}\BibitemShut {NoStop}%
\bibitem [{\citenamefont {Liu}\ \emph {et~al.}(2022)\citenamefont {Liu},
  \citenamefont {Gao}, \citenamefont {Wang}, \citenamefont {Xi}, \citenamefont
  {Hu}, \citenamefont {Wang}, \citenamefont {Liu}, \citenamefont {Lin},
  \citenamefont {Deng}, \citenamefont {Yang}, \citenamefont {Zhou},
  \citenamefont {Yang}, \citenamefont {Chong},\ and\ \citenamefont
  {Zhang}}]{Liu2022}%
  \BibitemOpen
  \bibfield  {author} {\bibinfo {author} {\bibfnamefont {G.-G.}\ \bibnamefont
  {Liu}}, \bibinfo {author} {\bibfnamefont {Z.}~\bibnamefont {Gao}}, \bibinfo
  {author} {\bibfnamefont {Q.}~\bibnamefont {Wang}}, \bibinfo {author}
  {\bibfnamefont {X.}~\bibnamefont {Xi}}, \bibinfo {author} {\bibfnamefont
  {Y.-H.}\ \bibnamefont {Hu}}, \bibinfo {author} {\bibfnamefont
  {M.}~\bibnamefont {Wang}}, \bibinfo {author} {\bibfnamefont {C.}~\bibnamefont
  {Liu}}, \bibinfo {author} {\bibfnamefont {X.}~\bibnamefont {Lin}}, \bibinfo
  {author} {\bibfnamefont {L.}~\bibnamefont {Deng}}, \bibinfo {author}
  {\bibfnamefont {S.~A.}\ \bibnamefont {Yang}}, \bibinfo {author}
  {\bibfnamefont {P.}~\bibnamefont {Zhou}}, \bibinfo {author} {\bibfnamefont
  {Y.}~\bibnamefont {Yang}}, \bibinfo {author} {\bibfnamefont {Y.}~\bibnamefont
  {Chong}},\ and\ \bibinfo {author} {\bibfnamefont {B.}~\bibnamefont {Zhang}},\
  }\bibfield  {title} {\bibinfo {title} {{Topological Chern vectors in
  three-dimensional photonic crystals}},\ }\href
  {https://doi.org/10.1038/s41586-022-05077-2} {\bibfield  {journal} {\bibinfo
  {journal} {Nature}\ }\textbf {\bibinfo {volume} {609}},\ \bibinfo {pages}
  {925} (\bibinfo {year} {2022})}\BibitemShut {NoStop}%
\bibitem [{\citenamefont {Zhou}\ \emph {et~al.}(2023)\citenamefont {Zhou},
  \citenamefont {Lai}, \citenamefont {Xie}, \citenamefont {Sun}, \citenamefont
  {He},\ and\ \citenamefont {Chen}}]{PhysRevB.107.014105}%
  \BibitemOpen
  \bibfield  {author} {\bibinfo {author} {\bibfnamefont {Y.-C.}\ \bibnamefont
  {Zhou}}, \bibinfo {author} {\bibfnamefont {H.-S.}\ \bibnamefont {Lai}},
  \bibinfo {author} {\bibfnamefont {J.-L.}\ \bibnamefont {Xie}}, \bibinfo
  {author} {\bibfnamefont {X.-C.}\ \bibnamefont {Sun}}, \bibinfo {author}
  {\bibfnamefont {C.}~\bibnamefont {He}},\ and\ \bibinfo {author}
  {\bibfnamefont {Y.-F.}\ \bibnamefont {Chen}},\ }\bibfield  {title} {\bibinfo
  {title} {Magnetic corner states in a two-dimensional gyromagnetic photonic
  crystal},\ }\href {https://doi.org/10.1103/PhysRevB.107.014105} {\bibfield
  {journal} {\bibinfo  {journal} {Phys. Rev. B}\ }\textbf {\bibinfo {volume}
  {107}},\ \bibinfo {pages} {014105} (\bibinfo {year} {2023})}\BibitemShut
  {NoStop}%
\bibitem [{\citenamefont {Zhou}\ \emph {et~al.}(2024)\citenamefont {Zhou},
  \citenamefont {Liu}, \citenamefont {Wang}, \citenamefont {Li}, \citenamefont
  {Xie}, \citenamefont {Zhang}, \citenamefont {Mandal}, \citenamefont {Xi},
  \citenamefont {Gao}, \citenamefont {Deng},\ and\ \citenamefont
  {Zhang}}]{10.1093/nsr/nwae121}%
  \BibitemOpen
  \bibfield  {author} {\bibinfo {author} {\bibfnamefont {P.}~\bibnamefont
  {Zhou}}, \bibinfo {author} {\bibfnamefont {G.-G.}\ \bibnamefont {Liu}},
  \bibinfo {author} {\bibfnamefont {Z.}~\bibnamefont {Wang}}, \bibinfo {author}
  {\bibfnamefont {S.}~\bibnamefont {Li}}, \bibinfo {author} {\bibfnamefont
  {Q.}~\bibnamefont {Xie}}, \bibinfo {author} {\bibfnamefont {Y.}~\bibnamefont
  {Zhang}}, \bibinfo {author} {\bibfnamefont {S.}~\bibnamefont {Mandal}},
  \bibinfo {author} {\bibfnamefont {X.}~\bibnamefont {Xi}}, \bibinfo {author}
  {\bibfnamefont {Z.}~\bibnamefont {Gao}}, \bibinfo {author} {\bibfnamefont
  {L.}~\bibnamefont {Deng}},\ and\ \bibinfo {author} {\bibfnamefont
  {B.}~\bibnamefont {Zhang}},\ }\bibfield  {title} {\bibinfo {title}
  {{Realization of a quadrupole topological insulator phase in a gyromagnetic
  photonic crystal}},\ }\href {https://doi.org/10.1093/nsr/nwae121} {\bibfield
  {journal} {\bibinfo  {journal} {National Science Review}\ ,\ \bibinfo {pages}
  {nwae121}} (\bibinfo {year} {2024})}\BibitemShut {NoStop}%
\bibitem [{\citenamefont {Jackiw}\ and\ \citenamefont
  {Rebbi}(1976)}]{PhysRevD.13.3398}%
  \BibitemOpen
  \bibfield  {author} {\bibinfo {author} {\bibfnamefont {R.}~\bibnamefont
  {Jackiw}}\ and\ \bibinfo {author} {\bibfnamefont {C.}~\bibnamefont {Rebbi}},\
  }\bibfield  {title} {\bibinfo {title} {Solitons with fermion number
  \textonehalf{}},\ }\href {https://doi.org/10.1103/PhysRevD.13.3398}
  {\bibfield  {journal} {\bibinfo  {journal} {Phys. Rev. D}\ }\textbf {\bibinfo
  {volume} {13}},\ \bibinfo {pages} {3398} (\bibinfo {year}
  {1976})}\BibitemShut {NoStop}%
\bibitem [{\citenamefont {Haldane}(1988)}]{PhysRevLett.61.2015}%
  \BibitemOpen
  \bibfield  {author} {\bibinfo {author} {\bibfnamefont {F.~D.~M.}\
  \bibnamefont {Haldane}},\ }\bibfield  {title} {\bibinfo {title} {Model for a
  quantum hall effect without landau levels: Condensed-matter realization of
  the "parity anomaly"},\ }\href {https://doi.org/10.1103/PhysRevLett.61.2015}
  {\bibfield  {journal} {\bibinfo  {journal} {Phys. Rev. Lett.}\ }\textbf
  {\bibinfo {volume} {61}},\ \bibinfo {pages} {2015} (\bibinfo {year}
  {1988})}\BibitemShut {NoStop}%
\bibitem [{\citenamefont {Pozar}()}]{Pozar}%
  \BibitemOpen
  \bibfield  {author} {\bibinfo {author} {\bibfnamefont {D.~M.}\ \bibnamefont
  {Pozar}},\ }\href {https://search.library.wisc.edu/catalog/9910153599402121}
  {\emph {\bibinfo {title} {{Microwave engineering}}}}\ (\bibinfo  {publisher}
  {Fourth edition. Hoboken, NJ : Wiley, [2012]
  {\textcopyright}2012})\BibitemShut {NoStop}%
\end{thebibliography}%

\vspace{15pt}
\noindent \textbf{\large{Acknowledgments}}
\vspace{5pt}

\noindent  This work is supported by the Singapore National Research Foundation Competitive Research Program (grant no. NRF-CRP23-2019-0007), Singapore Ministry of Education Academic Research Fund Tier 2 (grant no. MOE-T2EP50123-0007) and Tier 1 RG81/23. The work at Southern University of Science and Technology was sponsored by the National Natural Science Foundation of China (No. 62375118, No 6231101016, and No. 12104211), Guangdong Basic and Applied Basic Research Foundation (No. 2024A1515012770), and Shenzhen Science and Technology Innovation Commission (No. 20220815111105001).

\vspace{15pt}
\noindent \textbf{\large{Authors Contributions}}
\vspace{5pt}

\noindent  S.M. and G.-G.L. initiated the idea and designed the structure. S.M. provided the theory with the help of R.B. and H.T.T. G.-G.L., S.M., and Z.W. performed the simulation. Z.G., X.X. and Z.W. fabricated samples. Z.W., X.X., Z.G. and P.Z. carried out the measurements. S.M. wrote the manuscript with the help of G.-G.L., Z.W., Z.G., and B.Z. B.Z., G.-G.L., Z. G., and X.X. supervised the project.

\vspace{15pt}
\noindent \textbf{\large{Competing Interests}}
\vspace{5pt}

\noindent The authors declare no competing interests. 

\end{document}